\documentclass[a4paper,11pt]{article}
\pdfoutput=1 

\usepackage{jcappub} 

\usepackage[T1]{fontenc} 

\usepackage{cleveref}
\usepackage{natbib}
\usepackage{amssymb}
\usepackage{amsmath}
\usepackage{siunitx}
\usepackage{physics}
\usepackage{graphicx,xcolor}
\usepackage{booktabs}
\usepackage{bm}
\usepackage{multirow}
\usepackage{fontawesome}

\newcommand{\cM}{\mathcal{M}}
\newcommand{\cJ}{\mathcal{J}}
\newcommand{\cH}{\mathcal{H}}

\usepackage{verbatim}

\newcommand{\HazmaOne}{{\itshape\bfseries Hazma 1}}
\newcommand{\HazmaTwo}{{\itshape\bfseries Hazma 2}}
\newcommand{\Herwig}{{\itshape\bfseries Herwig}}
\newcommand{\HerwigDM}{{\itshape\bfseries Herwig4DM}}

\title{\huge Hazma Meets HERWIG4DM:\\ Precision Gamma-Ray, Neutrino, and Positron Spectra for Light Dark Matter}

\affiliation[a]{Ciela -- Computation and Astrophysical Data Analysis Institute, Montréal, Quebec, Canada}
\affiliation[b]{Département de Physique, Université de Montréal, 1375 Avenue Thérèse-Lavoie-Roux, Montréal, QC H2V 0B3, Canada}
\affiliation[c]{Mila -- Quebec AI Institute, 6666 St-Urbain, \#200, Montreal, QC, H2S 3H1}
\affiliation[d]{Institut f\"ur Theoretische Physik, Universit\"at Heidelberg, Germany}
\affiliation[e]{Department of Physics and Santa Cruz Institute for Particle Physics, University of California, Santa Cruz, 1156 High Street, Santa Cruz, CA 95064, USA}
\affiliation[f]{Instituto de F\'isica, Universidade de S\~ao Paulo, 05508-090 S\~ao Paulo, SP, Brasil}

\author[a,b,c]{Adam Coogan,}
\author[e]{Logan Morrison,}
\author[d]{Tilman Plehn,}
\author[e]{Stefano Profumo,}
\author[f]{Peter Reimitz}

\emailAdd{adam.coogan@umontreal.ca}
\emailAdd{loanmorr@ucsc.edu}
\emailAdd{profumo@ucsc.edu}
\emailAdd{plehn@uni-heidelberg.de}
\emailAdd{peter@if.usp.br}

\abstract{
    We present a new open-source package, \HazmaTwo, that computes accurate spectra relevant for indirect dark matter searches for photon, neutrino, and positron production from vector-mediated dark matter annihilation and for spin-one dark matter decay. The tool bridges across the regimes of validity of two state of the art codes: \HazmaOne, which provides an accurate description below hadronic resonances up to center-of-mass energies around 250 MeV, and \HerwigDM, which is based on vector meson dominance and measured form factors, and accurate well into the few GeV range. The applicability of the combined code extends to approximately 1.5 GeV, above which the number of final state hadrons off of which we individually compute the photon, neutrino, and positron yield grows exceedingly rapidly. We provide example branching ratios, particle spectra and conservative observational constraints from existing gamma-ray data for the well-motivated cases of decaying dark photon dark matter and vector-mediated fermionic dark matter annihilation. Finally, we compare our results to other existing codes at the boundaries of their respective ranges of applicability. \HazmaTwo\ is freely available on GitHub \href{https://github.com/LoganAMorrison/Hazma}{\faGithub}.
}

\begin{document}
\maketitle
\flushbottom

\clearpage
\section{Introduction}
\label{sec:introduction}
The hunt for non-gravitational signals from the dark matter sector has so far yielded null results, although admittedly intriguing possible hints of a signal have emerged over the years (see e.g. Ref.~\cite{ParticleDataGroup:2020ssz} for a review). The paradigm of weakly-interacting massive particle (WIMP) dark matter models - predicated on the notion that the dark matter participates in the Standard Model weak interactions and that the production of dark matter in the early universe is via thermal freeze-out - has motivated a strong focus on electroweak scale annihilation products. Theoretically this is largely, in turn, due to arguments based on the classic result of Lee and Weinberg \cite{Lee:1977ua}, indicating that thermal WIMPs lighter than, roughly, a few GeV would inevitably overclose the universe in a standard cosmological scenario. Experimentally, GeV gamma-ray telescopes and cosmic-ray detectors such as the Fermi Large Area Telescope and its predecessors COMPTEL and EGRET, and PAMELA, AMS-02 and CALET have provided numerous observations used to set constraints on dark matter models with masses in the few GeV up to several-TeV range \cite{Profumo:2019ujg}.

As a result of both the lack of observational facilities, and of signals from dark matter annihilation and decay in the GeV range, in recent years the MeV range has come to the forefront of both the theoretical and observational landscape of dark matter searches (e.g. \cite{Coogan:2019qpu} and references therein). This also holds for direct dark matter searches \cite{Mitridate:2022tnv}. Since COMPTEL and, in part, EGRET observations, the MeV gamma-ray sky has largely remained unexplored. While a departure from the WIMP paradigm is required to postulate a dark matter mass in the MeV, and while rather strong constraints can be placed in certain cases on the annihilation rate of MeV-scale dark matter from cosmology \cite{Lehmann:2020lcv}, exploring the possible signals of MeV dark matter models is a very timely endeavor. First and foremost, due to the mentioned lack of observational facilities, numerous upcoming and concept telescopes are in the development stages, or under lively discussion. These include, but are not limited to, e.g., All-Sky ASTROGAM \cite{e_astrogam, as_astrogam}, GECCO \cite{Orlando:2021get,Coogan:2021rez}, AdEPT~\cite{adept}, AMEGO~\cite{amego}, MAST~\cite{mast}, COSI~\cite{Tomsick}, PANGU~\cite{pangu} and GRAMS~\cite{grams}.

On the theory side, it has become apparent that the passel of available codes for the calculation of the photon and positron yield from dark matter annihilation or decay in the GeV is inapplicable below the GeV scale, where the relevant degrees of freedom are not partons but hadronic resonances. These codes include PPPC4DMID \cite{Cirelli:2010xx}, micrOMEGAs \cite{Belanger:2008sj}, MadDM \cite{Arina:2021gfn}, and DarkSUSY \cite{Bringmann:2018lay}, which utilize a description based on hard perturbative processes followed by parton showers, fragmentation, and decay of hadronic products, and are in turn built via Monte Carlo simulation tools such as \Herwig\ \cite{Bellm:2019zci} and Pythia \cite{Sjostrand:2019zhc} (see~\cite{Amoroso:2018qga} for a uncertainty estimate thereof). For this reason, a subset of the authors have implemented an extension to \Herwig\ called \HerwigDM, which utilizes the vector meson dominance (VMD) framework supplemented with data from $e^+e^-$ collisions \cite{Plehn:2019jeo, Reimitz:2021wcq}. A second subset of the authors developed a distinct package, \HazmaOne\ \cite{Coogan:2019qpu}, that treats accurately the decay products of hadrons and, generally, the annihilation or decay of sub-GeV dark matter to hadrons in the context of chiral perturbation theory (ChPT) \cite{Coogan:2021sjs}.

In this work we merge the two codes \HerwigDM\ and \HazmaOne\ into an update of the latter code called \HazmaTwo, which provides a complete framework to describe dark matter annihilation via a vector mediator, or the decay of vector (spin 1) dark matter for any mass below $\sim1.5$ GeV. The upper limit is constrained by the untreatable large number of hadronic final states that can be produced as a result of hadronization of quark-antiquark pairs from dark matter annihilation or decay, and whose photon and positron decay yields would need to be individually implemented in \HazmaOne. At about 1.5 GeV the existing codes listed above provide an increasingly reliable output, and begin to match the results of \HazmaTwo, as we discuss in detail below.

The remainder of this study has the following structure: In the next section we describe the \HazmaOne\ and \HerwigDM\ codes, their theoretical underpinnings, and details on the transition regime between the ranges of applicability of the two codes. The following \cref{sec:spectra} describes the calculation of meson decays, shows outputs of the code in the form of branching ratios for a couple illustrative dark matter models, and the resulting photon and positron spectra. There we additionally derive conservative model constraints from existing gamma-ray data. \Cref{sec:discussion} presents our discussion and conclusions, and the two Appendices elaborate on issues related to form factors and decay spectra.

Our code is publicly available on GitHub at \url{https://github.com/LoganAMorrison/Hazma}.

\section{Hazma and HERWIG4DM}
When looking for dark matter below the typical WIMP scale, the most interesting mass scale affecting annihilating dark matter patterns is at or below the GeV scale. Below the confinement scale of QCD, strong interactions are described by ChPT, where pseudoscalar and vector mesons are the relevant hadronic degrees of freedom instead of partons \cite{Gasser:1984gg, Weinberg:1978kz,Scherer:2002tk}. The corresponding effective field theory (EFT) describes the dynamics of pseudoscalar mesons such as pions and kaons that are treated as pseudo-Goldstone bosons under the chiral symmetry group $SU(3)_L\times SU(3)_R$. This symmetry can be enhanced to a ``hidden'' local gauge theory with vector fields acting as gauge fields of a certain flavor, as seen in the following (see also \cite{Bando:1984ej,Kramer:1984cy,Bando:1987br,Bando:1985rf}). Fields that do not transform under QCD, like the photon field or beyond-the-standard model vector fields can be introduced as external fields linked to a $U(1)$ symmetry, which can be understood as a subgroup of the local $SU(3)_L\times SU(3)_R$ symmetry. The BSM field that we will introduce in the following, and use throughout the paper and in the \HazmaTwo~code up to 1.5 GeV is a vector particle with equal couplings to left- and right-handed quarks.

At slightly higher masses as the applicability of ChPT, the hadronic $\rho$, $\omega$. and $\phi$ resonances of QCD are best described by the VMD model~\cite{Sakurai:1960ju,Kroll:1967it,Lee:1967iv}. VMD parameters can be extracted from $e^+e^-$ data through a set of hadronic currents \cite{Ezhela:2003pp,ParticleDataGroup:2020ssz}. We connect  these two regimes  using the results of Refs.~\cite{Coogan:2019qpu,Coogan:2021sjs} for the ChPT and Refs.~\cite{Plehn:2019jeo,Foguel:2022ppx} for VMD, and focus on a smooth and consistent transition between the two regimes.

\paragraph{ChPT.} Using the notation of Ref.~\cite{Coogan:2021sjs}, the
pseudoscalar states can be represented as the matrix,
\begin{align}
  \Phi(x)=\sum_{a=1}^8 \lambda^a\phi^a=
  \begin{pmatrix}
    \pi^0+\frac{1}{\sqrt{3}}\eta & \sqrt{2}\pi^+                 & \sqrt{2} K^+ \\
    \sqrt{2} \pi^-               & -\pi^0+\frac{1}{\sqrt{3}}\eta & \sqrt{2}K^0  \\
    \sqrt{2}K^-                  & \sqrt{2}\bar{K}^0             &
    -\frac{2}{\sqrt{3}}\eta\end{pmatrix} \; ,
  \label{eq:Phi}
\end{align}
with pions $\pi^0, \pi^\pm$, kaons $K^0, K^\pm$ and the eta meson $\eta$, and
where $\lambda_a$ are the Gell-Mann matrices with the normalization
$\text{Tr}(\lambda^a \lambda^b)=2 \delta^{ab}$. In terms of the
Goldstone matrix
\begin{align}
  \Sigma=\exp \frac{i\Phi}{f_\pi}
\end{align}
and the pion decay constant $f_\pi$ the leading-order ChPT Lagrangian with
chiral symmetry is
\begin{align}
  \mathcal{L}_{\rm ChPT}=\frac{f_\pi^2}{4}{\rm Tr}\left[(D_\mu
    \Sigma)(D^\mu \Sigma)^\dagger\right]+\frac{f_\pi^2}{4}{\rm
    Tr}\left[\chi^\dagger\Sigma+\Sigma^\dagger \chi\right] \; .
  \label{eq:LChPT}
\end{align}
The covariant derivative $D_\mu$ acting on the Goldstone matrix is $D_\mu \Sigma
  = \partial_\mu \Sigma - i r_\mu\Sigma + i\Sigma l_\mu$, where $r_\mu$ and
$l_\mu$ are right-handed and left-handed chiral currents, respectively.

In addition to the photon field $A_\mu$, we can incorporate a new massive vector boson $V_\mu$ into the chiral Lagrangian. The vector's interactions with the fermionic DM field $\chi$, the quarks and the photon are given by
\begin{align}
  \mathcal{L}_{{\rm Int}(V)}=V_\mu\left( g_{V\chi} \, \bar{\chi}\gamma^\mu\chi + \bar{q} \, G_{Vq} \, \gamma^\mu q \right)-\frac{\epsilon}{2}V^{\mu\nu}F_{\mu\nu} \; .
  \label{eq:Xint}
\end{align}
The quark-mediator coupling matrix is given by $G_{Vq}=\text{diag}(g_{Vu},g_{Vd},g_{Vs})$.

In the following and regarding the applicability of the \HazmaTwo~extension towards 1.5~GeV, we describe the case of a vector particle coupling equally to left-handed and right-handed quarks. In terms of ChPT, this means that we add a new vector field $V_\mu$ to the chiral Lagrangian with $r_\mu=l_\mu= G_{Vq} V_\mu-eQ^{\rm em}A_\mu$ and $Q^{\rm em}=\text{diag}(q_u,q_d,q_s)=\text{diag}(2/3,-1/3,-1/3)$. In general, other coupling structures could be considered as for example unequal couplings to left- and right-handed fields as for the charged W-boson in the SM with $r_\mu=0$, and $l_\mu=-\frac{g}{\sqrt{2}}(W_\mu^+ T_+ + {\rm h.c.})$ as well as other boson types like axial-vectors, pseudoscalars or scalars. While we keep the \HazmaOne~description for both spin-0 and spin-1 for energies in the range of ChPT ($\sqrt{s}\lesssim 500$~MeV), we focus on spin-1 particles with equal couplings to left- and right-handed quarks in the extension to \HazmaTwo~for energies up to 1.5~GeV and leave the description of other couplings and spin configurations in the extended energy range for future work. 

\paragraph{VMD model.} QCD vector mesons such as the $\rho$, $\omega$, and $\phi$ are considered composite fields identified with dynamical gauge bosons of a hidden local symmetry in the chiral Lagrangian. We introduce the non-Abelian vector gauge fields $X_\mu \equiv X_\mu^a \lambda^a/2=X_\mu^a T^a$ with group generators $T^a$ and include the vector mesons in the covariant derivative. Linear combinations of those gauge fields and group generators will then describe the physical vector mesons as seen below in eq.~\ref{eq:generators}. Expanding out the covariant derivative gives interactions of the form
\begin{align}
  \mathcal{L}_{{\rm Int}(X)}=-\frac12 g_{XPP} \, \text{Tr} \left[X_\mu,\left[\Phi,\partial^\mu \Phi \right] \right]+ 2 g_{X\gamma} \; A_\mu \text{Tr}[X^\mu Q^\text{em}]+...
  \label{eq:XPP}
\end{align}
The Lagrangian above includes the vector-pseudoscalar coupling $g_{XPP}$ and a photon-vector mixing.

Following the standard procedure of ChPT, we have to add Wess-Zumino-Witten (WZW) terms~\cite{Wess:1971yu,Witten:1983tw} to consistently describe the theory of mesons~\cite{Fujiwara:1984mp,Kramer:1984cy,Bando:1987br}. In principle, there are several non-vanishing WZW terms that could describe the dynamics of vector and pseudoscalar mesons, and only experimental measurements tell us which linear combination of WZW terms survive and describe the theory accurately. The interaction term for vector mesons is given by
\begin{align}
  \mathcal{L}_{XX\Phi} & =g_{XXP}\epsilon^{\mu\nu\rho\sigma}\,  \partial_\mu X_\nu^a \, \partial_\rho X_\sigma^b \, \phi^c \; {\rm Tr}\left[T^a T^b T^c\right].
  \label{eq:XXP}
\end{align}
%
If we now introduce the vector mediator $V_\mu$ for the given DM model, we find a
mediator-meson mixing term of the form
\begin{align}
  \mathcal{L}_{{\rm Int}(XV)}=2 g_{XV} \, V_\mu {\rm Tr}\left[X^\mu G_{Vq}\right] \; .
  \label{eq:VX}
\end{align}
A typical process in the VM-dominated regime could then look like
\begin{align}
  V\to X \to X'\pi \to 3\pi \; ,
  \label{eq:mix_dec}
\end{align}
where the mediator is mixing with a vector meson $X$, which decays to another
vector meson and a pion, followed by the decay $X'\to 2\pi$. In this example,
the decay $X\to 3\pi$ is fully described within the Standard Model (SM). In general, once the
vector mediator is converted into a vector meson, the hadronic meson decays
determine the rest of the decay chain. This is why we can use $e^+e^-$
annihilation data to fit the hadronic currents $J_\mathcal{H}^\mu$ precisely.
All we need to do is change the matrix element of the process by replacing the
leptonic current $L_\mu$ with the polarization vector of the mediator
$\varepsilon^*(V)$~\cite{Foguel:2022ppx}
\begin{align}
  \mathcal{M}_{e^+e^-\to \, \mathcal{H}}
   & =L_\mu J^\mu_\mathcal{H}
   & \Rightarrow \qquad
  \mathcal{M}_{V\to \mathcal{H}}
   & =\varepsilon_{\mu}(V) \sum_X
  \frac{ {\rm Tr}\left[T_X G_{Vq}\right]}{{\rm Tr}\left[T_X Q^{\rm em}\right]} \;  J^\mu_{\mathcal{H}}(X)  \; ,
  \label{eq:ME}
\end{align}
with the appropriate rescaling of the photon-meson coupling to the
mediator-meson coupling for a vector meson resonance $X=\rho,\omega,\phi$. The
generators $T_X$ are given by
\begin{align}
  T_\rho = \frac12 \text{diag}(1,-1,0) \qquad \quad
  T_\omega = \frac12 \text{diag}(1,1,0)\qquad \quad
  T_\phi = \frac12 \text{diag}(0,0,1) \; .
  \label{eq:generators}
\end{align}
For a detailed description of all possible hadronic currents including their
form factors we refer to Ref.~\cite{Foguel:2022ppx}. In the mediator mass range
up to 1.5~GeV\footnote{For more details see Fig.~1 and~2 in
  Ref.~\cite{Foguel:2022ppx}}, the dominant hadronic channels are $\pi\gamma,
  2\pi, 3\pi, 4\pi, KK, KK\pi, \eta\gamma, \eta\pi\pi, \omega\pi, \omega\pi\pi,
  \phi\pi, \eta\omega$, and $\eta\phi$. Using its similarity to the $\eta\pi\pi$
channel, we also include the $\eta'\pi\pi$ channel.

\paragraph{Transition regime.} Linking the ChPT and VMD descriptions we have to
ensure that the effects of resonant vector mesons vanish in the low-energy limit
$\hat{s}\ll m_{\omega,\rho}$. We require specifically $F(0)\to 1$ for
$\hat{s}\to 0$ for the VMD form factors $F(\hat{s})$ in the hadronic current,
for all leading-order ChPT contributions. This guarantees that the vector
mediator directly couples to the final states in the low energy limit, whereas
in the hadronic resonance region it first mixes and then decays as in
\cref{eq:mix_dec}. The hadronic channels considered in the \HazmaOne\ code \cite{Coogan:2019qpu} and
well described by ChPT are  $V \to \pi^+\pi^-$ and $V\to \pi^0\gamma$. The
$\pi^+\pi^-$ channel is fixed by gauge invariance, and the transition is given
by
\begin{align}
  \eval{J_\mathcal{\pi\pi}^\mu}_\text{VMD} = -(p_1-p_2)^\mu \, F_{\pi\pi}(\hat{s})
  \quad \rightarrow \quad
  \eval{J_\mathcal{\pi\pi}^\mu}_\text{ChPT} = -(p_1-p_2)^\mu \; ,
\end{align}
with the pion form-factor $F_{\pi\pi}(\hat{s})$ from Ref.~\cite{Czyz:2010hj}.

The $\pi^0\gamma$ channel is described by an anomalous WZW term. We parametrize
the hadronic current based on Ref.~\cite{Czyz:2017veo}, but in a simplified form
as
\begin{equation}
\begin{split}
  \eval{J_{\pi\gamma}^\mu}_\text{VMD} & =\varepsilon^{\mu\nu\rho\sigma}q_\nu \varepsilon_{\rho,\gamma}p_{\sigma,\gamma} \, F_{\pi\gamma}(q^2), \\
  \text{with}\quad F_{\pi\gamma}(q^2) & =-\frac{e}{4\pi^2f_\pi}+a\hat{s}\frac{4\sqrt{2}e}{3f_\pi}\sum_{X=\rho,\omega,\phi} \frac{a_X}{q^2-m_X^2+im_X\Gamma_X} \; ,
\end{split}
\label{eq:pigamma}
\end{equation}
where $q^\mu=p^\mu_{\pi}+p^\mu_{\gamma}$ is the total momentum for the final
states. The form factor contains a direct anomalous coupling of the vector boson
to $\pi\gamma$, in addition to one including vector mesons as resonant states
and which vanishes as $\hat{s}\to 0$. This perfectly matches the ChPT form
\begin{align}
  \eval{J^\mu_{\pi\gamma}}_\text{ChPT} =-\frac{e}{4\pi^2f_\pi}\varepsilon^{\mu\nu\rho\sigma}q_\nu \varepsilon_{\rho,\gamma}p_{\sigma,\gamma} \; ,
  \label{eq:JChPT}
\end{align}
that can be obtained from the Lagrangian as described in
Ref.~\cite{Coogan:2019qpu}. While the vector meson dominated part can be
rescaled through coupling ratios like the one given in \cref{eq:ME}, to
account for arbitrary gauge boson couplings, the direct vector boson coupling to $\pi\gamma$ is rescaled by $(2g_{V_u}+g_{V_d})/(2q_u+q_d)=2g_{V_u}+g_{V_d}$.
Details of the fit are presented in Appendix~\ref{app:formfactor}.

\paragraph{Form factors.} For illustration purposes, we sketch the general
approach to describe form factors in the VMD model. The vector meson resonances
$\rho, \omega$, and $\phi$ are the most essential contributors to the form factor. Which
of those resonances is dominant for a given channel can be derived from the
interaction Lagrangians in \cref{eq:XPP} and \cref{eq:XXP} with the
$\rho\omega\pi$ and $\rho\phi\pi$ vertices being the most important one to
calculate subsequent processes for mesons. Alternatively, we can make use of
$G$-parity conservation for so-called light unflavoured mesons
($S=C=B=0$)~\cite{ParticleDataGroup:2020ssz}, which we can apply for pseudoscalar pions ($G=-1$), etas ($G=+1$), and the vector mesons $\rho$ ($G=+1$), $\omega$ ($G=-1$), and $\phi$ ($G=-1$). Since $G$-parity is an approximate multiplicative quantum number, an even number of pions in the final states can be traced back to a $\rho$, whereas an odd number of pions hint towards $\omega$ and $\phi$, for example.

If the final states include a photon, the question of the dominating vector boson is not defined, so it should be assumed that all mesons contribute. Nevertheless, experimental data show, for example, that only the $\omega$ and $\phi$ mesons decay to $\pi\gamma$ at a sizable rate.
In the case of kaons in the final states, we can rely on isospin symmetry and treat all mesons as $SU(2)$ isospin states to calculate the relative contribution of the $\rho, \omega$ and
$\phi$ isospin resonances~\cite{Davier:2010nc}. Processes with more than two final states are often best described by a chain of subsequent two-body decays, as described in \cref{eq:mix_dec}, including $\rho\omega\pi$ and $\rho\phi\pi$ vertices as given in \cref{eq:XXP}. The $3\pi$ channel can be resolved into the processes $\omega,\phi \to \rho^{\pm} \pi^\mp \to \pi^\pm \pi^0 \pi^\mp$ and a small isospin-breaking contribution $\rho^0\to \omega \pi^0 \to  \pi^+\pi^-\pi^0$~\cite{Czyz:2005as}. Other examples for consecutive decays of mesons are $4\pi$ final states with many possible substructures~\cite{Czyz:2008kw}, such as $\rho\to \omega\pi \to 4\pi$, or $\rho\to \rho f_0\to 4\pi$ or $KK\pi$ with $\rho,\phi\to K K^*\to KK\pi$~\cite{Foguel:2022ppx}. For some final states, like $\omega\pi\pi$, data indicates but does not clearly show which intermediate resonances should be accounted for~\cite{BaBar:2007qju,BaBar:2018rkc}. In these cases, it is sufficient to assume a point-like interaction, \textit{i.e.} a $\omega'\omega\pi\pi$ vertex to describe the decay $\omega'\to \omega\pi\pi$~\cite{Foguel:2022ppx}. Ultimately,  fits to data determine the strength of each resonance, and subsequent contributions to each channel.

By carefully treating the $\pi\gamma$ and $\pi\pi$ channels both in the ChPT and the VMD ranges by using the same parameterization, we guarantee a smooth transition between both regimes. For all other channels listed above it is sufficient, if not even required, to rely on the VMD results only.

\section{Example Spectra}\label{sec:spectra}

\begin{table}[t!]
  \centering
  \begin{small}
    \begin{tabular}{cccc}
      \toprule
      & Meson               & Intermediate decay products                                                       & Final stable particles \\
      \midrule
      \multirow{6}{*}{\rotatebox[origin=c]{90}{Pseudoscalars}} & $\pi^\pm$           & $\mu^\pm\nu_\mu$                                                                  & $e^\pm {3\nu}$                                                         \\
      & $\pi^0$             & -                                                                                 & $2\gamma$                                                                     \\
      & $\eta$              & $2\gamma$~(39\%) / $3\pi^0$~(32\%) / $\pi^+\pi^-\pi^0$~(23\%)                     & $2\gamma$ / $6\gamma$ /  $2e^\pm 2\gamma{6\nu}$                       \\
      & $K^0_S$             & $\pi^+\pi^-$~(69\%) /  $2\pi^0$~(31\%)                                            & $2e^\pm {6\nu}$ / $4\gamma$                                           \\
      & $K^0_L$             & $\pi^\pm e^\mp \nu_e$~(41\%) / $\pi^\pm \mu^\mp \nu_\mu$~(27\%) / $3\pi^0$~(20\%) & $2e^\pm{4\nu}$ / $2e^\pm{6\nu}$ / $6\gamma$                   \\
      & $K^\pm$             & $\mu^\pm \nu_\mu$~(64\%) / $\pi^\pm \pi^0$~(21\%) / $\pi^\pm\pi^\pm\pi^\mp$~(6\%) & $e^\pm{3\nu}$ / $e^\pm2\gamma{3\nu}$ / $3e^\pm{9\nu}$ \\
      \midrule
      \multirow{3}{*}{\rotatebox[origin=c]{90}{Vectors}} & $\rho$              & $\pi^+\pi^-$                                                                      & See decays of                                                                 \\
      & $\omega$            & $\pi^+\pi^-\pi^0$~(89\%) / $\pi^0\gamma$~(8\%)                                    & pseudoscalar mesons                                                           \\
      & $\phi$              & $K^+K^-$~(49\%) / $K^0_L K^0_S$~(34\%)                                            & above                                                                         \\
      \bottomrule
    \end{tabular}
  \end{small}
  \caption{A list of the dominant decay modes for pseudoscalar and vector mesons and the produced stable particle that leave imprints on indirect detection measurements. If no branching ratio percentage is given, the decay into those states is almost 100\%.}
  \label{tab:mesondecays}
\end{table}

Here we describe the energy spectra of neutrinos, positrons and gamma rays produced in DM annihilations via light vector mediators in the sub-GeV to GeV mass range. In the case of gamma rays, final state radiation from charged particles dominates for low-energy photons in the spectrum. At higher energies, the shape of the spectrum is determined by meson decays and by monochromatic photons directly produced in the decay of the vector particle. The positron spectrum consists of line-like sub-spectra from decays $V\to e^+e^-$, as well as a continuum piece coming from the decay of unstable leptons and mesons. Neutrinos are mostly produced in charged pion decays and hence, the continuum part of the spectrum is build up by the same processes as in the case of the positron. In addition, we might have a spectral line coming from $V\to \nu\bar{\nu}$ depending on the chosen vector model.

In the following, we consider a kinetic mixing model and a $B-L$ model. For both cases, we present the branching ratios into various hadronic final state configurations in order to tell which processes are dominating at a certain center-of-mass energy. In general, a greater number of different production modes for neutrinos, positrons and photons yields a broader spectrum. All results presented here are implemented in \HazmaTwo.

\subsection{Branching Ratios and Model Dependence}

\begin{figure}[t]
  \includegraphics[width=\textwidth]{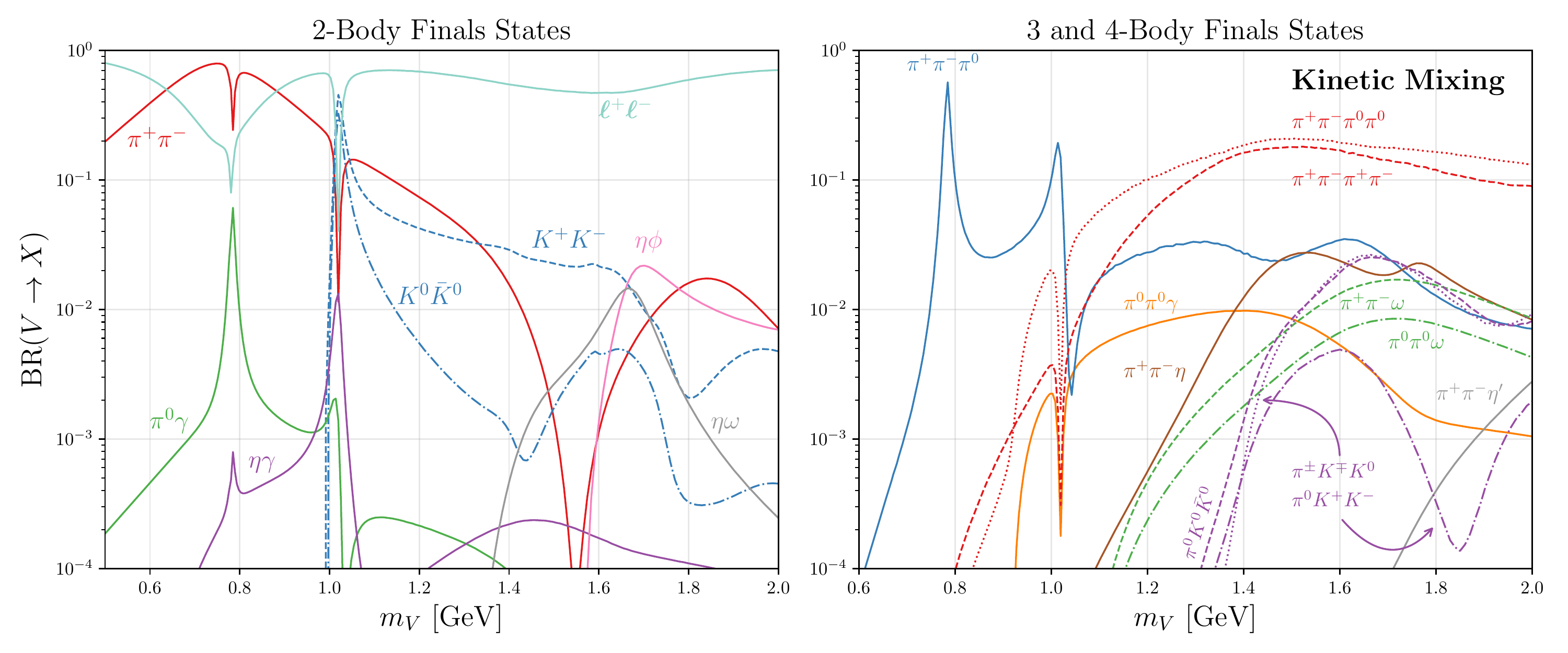}
  \caption{Branching ratios for a vector with kinetic-mixing couplings to the SM. For clarity, the panels highlight final states containing two (\textbf{left}) and three or four (\textbf{right}) particles.}
  \label{fig:BR_kinetic}
\end{figure}

\begin{figure}[t]
  \includegraphics[width=\textwidth]{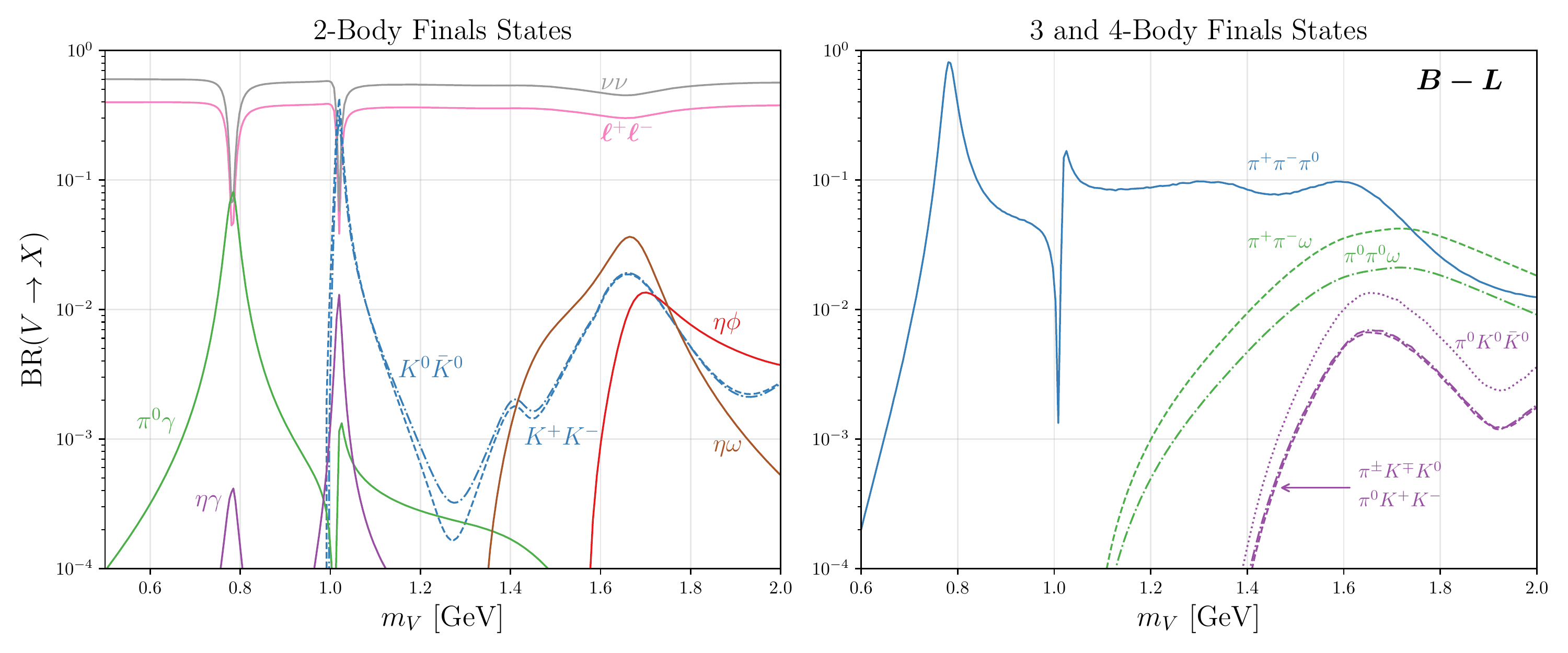}
  \caption{Branching ratios for a vector with $B-L$ couplings to the SM.}
  \label{fig:BR_BL}
\end{figure}

The vector mediator decay structure depends on how the couplings to SM fermions
are chosen in \cref{eq:Xint}. This determines if, and in which way, the
vector mediator is coupling to the vector mesons as seen from \cref{eq:ME}
and \cref{eq:generators}. In \cref{fig:BR_kinetic} and
\cref{fig:BR_BL}, we show two example models for
light vector mediators. For one, we choose the dark photon where only the
kinetic mixing with the SM photon in \cref{eq:Xint} fixes the coupling of the mediator
to SM fermions. In this case, the quark couplings are equal to the SM photon
case and the hadronic currents is entirely inherited from the SM photon
case. As a consequence, the branching fractions into two-, three- and four-body final states of the vector particle
should reflect the strength in which $e^+e^-\to~{\rm hadrons}$ are present in the SM. From \cref{fig:BR_kinetic}, we can clearly see that
multi-pion final states dominate the hadronic decay channels. The only exception
is the region around the $\phi$ resonance where two kaons in the final state
contribute as well. This implies that hadronic decays with predominantly
charged pions provide a positron source in the entire mass range considered
whereas we expect that the number of photons produced increases with mass/energy
when the kaon channels, as well as the $\pi^+\pi^-\pi^0\pi^0$ channel and
subdominant channels start playing a role. 

The situation changes drastically in the case
of the second model that we consider throughout the paper, the anomaly-free
$B-L$ model with gauge couplings to quarks and leptons. From the structure of
the quark-mediator coupling matrix $G_{Vq}=\text{diag}(1/3,1/3,1/3)$, we can
immediately see through \cref{eq:ME} and \cref{eq:generators} that the
$B-L$ gauge boson does not couple to the $\rho$ meson. Hence, all decay modes
that channel through the $\rho$ meson drop out as seen in \cref{fig:BR_BL}. Those are in particular, the $2\pi$ and $4\pi$ final states, as well as $\pi^0\pi^0\gamma, \pi^+\pi^-\eta$
and $\pi^+\pi^-\eta'$. As a consequence, we expect that for energy scales below $m_\phi\sim 1$~GeV positrons are only produced in direct production via direct $e^+e^-$ production or in decays of muons and pions in $\mu^+\mu^-$ and $\pi^+\pi^-\pi^0$, respectively. Photons can be produced via $\pi^0\gamma$, $\eta\gamma$, and $\pi^+\pi^-\pi^0$ and can be radiated of charged leptons below 1 GeV.

\subsection{Meson Decays}

Before we delve into the details of the spectrum calculations, we review the salient features of the main decay channels of light mesons (see also Ref.~\cite{Christy:2022pvy} for a recent discussion). The production of $\gamma$ and $e^\pm$ originates from many different meson decays, which we list in \cref{tab:mesondecays}. The situation is relatively simple for pions, since virtually all charged pions decay via $\pi^+ \to \mu^+ \nu_\mu \to e^+\nu_e\bar{\nu}_\mu\nu_\mu$ whereas neutral pions almost entirely decay into photons $\pi^0\to \gamma\gamma$. For the other pseudoscalar mesons, namely etas and kaons, several decay modes open up. All of them include pions with the exceptions of the purely leptonic $K^+\to\mu^+\nu_\mu$ decay. While the $\eta$ meson mostly decays to photons, either directly $\eta\to2\gamma$ or through $3\pi^0 \to 6\gamma$, kaons dominantly decay to charged pions and leptons, meaning that kaons are a major source of positrons.

To describe vector meson decays to stable particles, we have to read \cref{tab:mesondecays} from the bottom up and go through all possibilities of produced pseudoscalar mesons and their consecutive decays. The simplest case is the $\rho$ meson where the decay chain is given by $\rho \to \pi^+\pi^- \to 2\mu2\nu_\mu\to 2e6\nu$. In contrast to that, the $\omega$ and $\phi$ mesons decay to pions and kaons respectively, which subsequently decay.

Neutrinos can be found in all processes that produce positrons. For charged pion decays, lepton flavor conservations requires that three neutrinos are produced at the end of the decay chain. Hence, we can say that almost every positron produced in hadronic processes comes with three neutrinos. The only exception is the $K_L^0\to \pi^\pm e^\mp \nu_e$ decay which directly gives a positron-neutrino pair.

\subsection{Gamma-ray spectra}

As already outlined in Ref.~\cite{Coogan:2019qpu}, there are three photon sources for gamma-ray spectra:
\begin{enumerate}
  \item \textbf{Final State Radiation (FSR).}  The spectral shape for photons coming from FSR goes like ${\rm d}N/{\rm d}E_\gamma \propto E_\gamma^{-1}$ at low-energies, in accordance with Low's theorem~\cite{Low:1958sn,Burnett:1967km}. As a consequence, radiation turns out to be the major photon source in the low-energy part of the spectrum. There are two ways of calculating the spectra from FSR for hadronic states. We can use the ChPT Lagrangian including an internal bremsstrahlung diagram with a 4-point vertex term $V \pi^+ \pi^- \gamma$ as used in Ref.~\cite{Coogan:2019qpu}. In this work, we mainly focus on energy scales above the validity of ChPT where we pass over to a regime where the Altarelli-Parisi spectrum gives a decent approximation, as seen in fig.~6 of Ref.~\cite{Coogan:2019qpu}. For leptonic final states, we rely on the exact results of Ref.~\cite{Coogan:2019qpu}.
  \item \textbf{Decays of unstable particles.} Photons can also be produced in pseudoscalar meson decays $\pi^0, \eta \to\gamma\gamma$. Those $\pi^0$s and $\eta$s can either be the final state of a certain channel, more precisely the $\pi^0$ channels $\pi^0\gamma,\ \pi^+\pi^-\pi^0,\ \pi^+\pi^-2\pi^0,\ K^0K^0\pi^0,\ K^+K^-\pi^0,\ \pi^0\pi^0\gamma,\ \omega2\pi^0$ and $\phi\pi^0$ as well as the $\eta$ channels $\eta\gamma,\ \eta\pi^+\pi^-,\ \eta\omega,\ \eta\phi$, or a decay product of final states like in the consecutive decay of the K-short $K_S^0\to 2\pi^0\to 4\gamma$. In the rest frame of the decaying pseudoscalar meson $P=\pi^0,\eta$, the gamma-ray spectrum should give two lines exactly at $E_\gamma=m_P/2$. In the lab frame this turns into a characteristic box spectrum around half the pseudoscalar meson energy $E_\gamma=m_P\gamma/2=E_P/2$ where $\gamma=E_P/m_P$ is the Lorentz boost factor. For the case of $\pi^0$s and $\eta$s being part of two-body final states, in particular $\pi^0\gamma$ and $\eta\gamma$, the central energy of the box can be calculated easily. For pions and etas in higher multiplicity final state configurations, or at the end of a decay chain, the kinematics are more complicated and the spectra be computed numerically. 
  \item \textbf{Monochromatic Gamma Rays.} Spectral lines can only come from $\pi^0\gamma$ and $\eta\gamma$ for vector mediator models. Those monochromatic gamma-ray lines are exactly at $E_\gamma=(E_{\rm CM}^2-m_P^2)/2E_{\rm CM}$ where $E_{\rm CM}$ is the center of mass energy of the process and $P=\pi^0,\eta$ is the pseudoscalar meson. Hence, we expect the spectral line in the $\eta\gamma$ process to be slightly shifted towards lower energies compared to the $\pi^0\gamma$ channel due to the mass difference of both pseudoscalar mesons involved.
\end{enumerate}

\begin{figure}[t]
  \includegraphics[width=\textwidth]{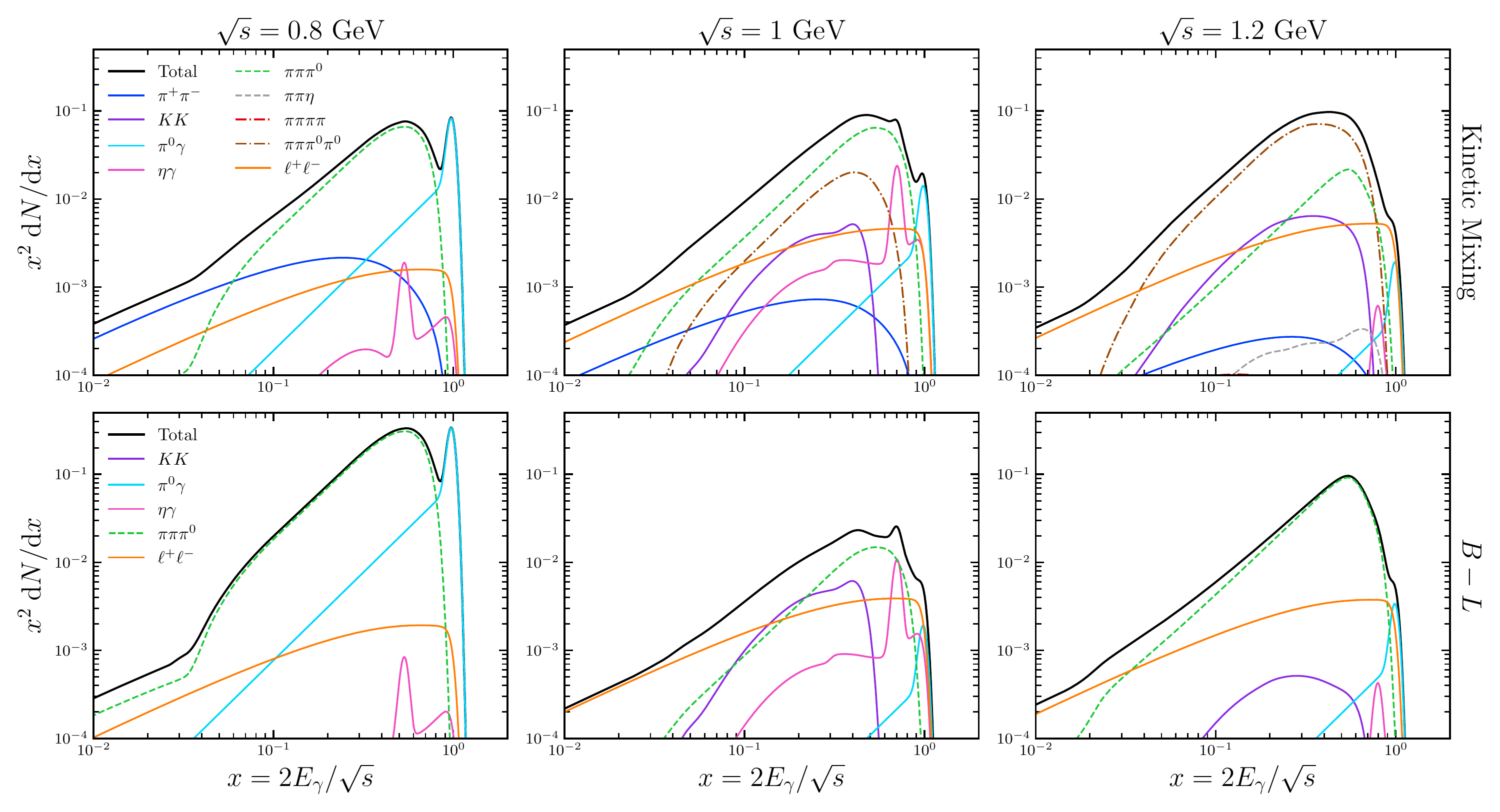}
  \caption{
    Photon spectrum for DM annihilations/vector decays with center-of-mass energies (\textbf{left}) 0.8 GeV, (\textbf{center}) 1 GeV and (\textbf{right}) 1.2 GeV. The rows show the spectra assuming (\textbf{top}) kinetic-mixing couplings and (\textbf{bottom}) $B-L$ couplings. In order to resolve \(\gamma\)-ray lines from \(\pi^{0}\gamma\) and \(\eta\gamma\), we have convolved the spectrum with a spectral energy distribution with an energy resolution of \(5\%\).
  }
  \label{fig:DNDX_PHOTON_KM_1GEV}
\end{figure}

To start off the discussion of all contributions from sub-channels to the total spectrum, let us first consider the kinetic mixing model (upper sequence of plots in \cref{fig:DNDX_PHOTON_KM_1GEV}); in the figure we show the gamma-ray spectra for center-of-mass energies of $\sqrt{s}=0.8, 1.0, 1.2$~GeV. The differences for the case of the $B-L$ model (lower sequence of panels of \cref{fig:DNDX_PHOTON_KM_1GEV}) will be detailed upon afterwards. An example for a gamma-ray spectrum purely coming from FSR is the $l^+l^-$ spectrum where we combined both the $e^+e^-$ and $\mu^+\mu^-$ spectra in all panels in \cref{fig:DNDX_PHOTON_KM_1GEV}. The leptonic decay channels are dominating through the whole mediator mass range as seen in \cref{fig:BR_kinetic} and hence, they contribute considerably to the total gamma-ray spectrum (black line) even though its not producing photons directly or in meson decays. Other examples are the purely FSR spectra for $\pi^+\pi^-$ and $2(\pi^+\pi^-)$. Whereas the $2(\pi^+\pi^-)$ mode is not a major source of photons, the $\pi^+\pi^-$ channel is the dominant FSR channel for energies around the $\rho$ mass $\sim 0.8$~GeV. For higher energies, the multi-charged-meson channels contribute less to the gamma spectrum due to smaller branching fractions and also drop down earlier the more heavy particles are involved in the process. For channels like $\pi^+\pi^-\pi^0$ and $K^+K^-$, the low-energy tail is coming from FSR as well, even though it is not visible in the given plot range of \cref{fig:DNDX_PHOTON_KM_1GEV}. 

Overall, the total gamma-ray spectrum is a sum of FSR contributions in the low-energy part of the spectrum. The $KK$ channel has a large BR around the $\phi$ mass $m_\phi\sim 1$~GeV and can, but not necessarily, contribute to the gamma-ray spectrum with photons through neutral pion decays in several different decay modes $K^0_S\to 2\pi^0, K^0_L\to 3\pi^0, K^\pm \to \pi^\pm \pi^0$ leading to a wavy accumulation of photons on top of the FSR spectrum. For the three-pion and four-pion final state the case is much clearer since $\pi^+\pi^-\pi^0$, and $\pi^+\pi^-\pi^0\pi^0$ guarantees two and four photons, respectively, coming from neutral pions. As a consequence, the $\pi^+\pi^-\pi^0$ channels is the leading contribution for the bulk part of the photon spectra up to 1~GeV, whereas $\pi^+\pi^-\pi^0\pi^0$ dominates above 1~GeV. The box spectrum from neutral pions in ${\rm d}N/{\rm d}E_\gamma$ translates to a linear shape in the given \cref{fig:DNDX_PHOTON_KM_1GEV}. This can clearly be seen in the $\pi^0\gamma$ channel which consists of a linear shape from $\pi^0\to \gamma\gamma$ decays and a spectral line at around $x=2E_\gamma/\sqrt{s}\approx 1$~GeV for all considered center-of-mass energies considered in \cref{fig:DNDX_PHOTON_KM_1GEV}. 

Whereas $\pi^0\gamma$ contributes to all considered spectra, it is more prominent for lower center-of-mass-energies. For the $\eta\gamma$ channel there is also a linear contribution to the spectrum from $\eta\to\gamma\gamma$. Nevertheless, it is accompanied by contributions from $\eta\to3\pi^0\to6\gamma$ and $\eta\to\pi^+\pi^-\pi^0$ and hence, the shape becomes round below the spectral line at $x=1-m_\eta^2/s\approx 0.7$ for $\sqrt{s}=1.0$~GeV. The imprint of the spectral lines on the final total energy spectrum for photons strongly depends on the energy resolution of the experiment. Notice that better instrumental energy resolution would produce a larger and more delta-like shaped peak for the spectral lines (in the figures we assumed an energy resolution $\Delta E/E=5\%$).

In \cref{fig:DNDX_RANGE}, we illustrate the dependence of the spectral shape with respect to the center-of-mass energy, studying  values between 200~MeV and 1.6 GeV in increments of 200 MeV with a slight change in color and alternating line style. At $\sqrt{s}=200$~MeV, the only available channels are $e^+e^-$, and with a small branching fraction, $\pi^0\gamma$. Hence, the total spectrum is mainly dominated by FSR photons going like ${\rm d}N/{\rm d}E_\gamma\propto E_\gamma^{-1}$ which translates to a characteristic increasing shape in our representation of the plot. There is not yet a visible effect of $\pi^0\gamma$ such as a tiny peak from the direct photon visible in the higher energy end of the spectrum where FSR drops sharply. For increased center-of-mass energy of $\sqrt{s}=400$~MeV and $\sqrt{s}=600$~MeV and a larger branching ratio into $\pi^0\gamma$, this peak is more pronounced and stands out in the total spectrum. We also notice a slightly bigger FSR contribution in the lower energy end coming from $\pi^+\pi^-$. Getting closer to the first resonances of the $\rho$ and $\omega$ around 800 MeV, photons from $\pi^0\to \gamma\gamma$ decays contribute just below $x<1$. In particular, contributions from the neutral pion in the $\pi\gamma$, and three-pion channel impact the total spectrum. 

Above 800 MeV the spectra continue to increase in the bulk part with additional four-pion and $\eta\gamma$ channels opening up and especially a more dominant role of $\pi^+\pi^-\pi^0$. Once the phase-space suppression of the $4\pi$ channels has settled, the $\pi^+\pi^-\pi^0\pi^0$ channel dominantly provides photons to the spectrum for higher $E_\gamma$. For further increasing center-of-mass energies well above $\sqrt{s} \gg m_{\pi^0,\eta}$, the effect of mesonic decay photons leading to a higher maximum of the total spectrum ceases with decreasing hadronic branching fractions. We would observe a broadening of the spectrum and a convergence towards a FSR-like spectral shape. If no new channels were added above 1.6~GeV, the spectrum would be completely dominated by FSR from the dominant lepton channels again. 

Between 1.6~GeV and 2 GeV, we also expect that the hadronic channels go over into a perturbative regime of $V\to q\bar{q}$ processes followed by a parton shower, hadronization and the decay of those hadrons. It is uncertain where exactly the regime of direct production of hadrons ends and the perturbative range starts. As seen in Ref.~\cite{Foguel:2022ppx}, above $\sqrt{s}>1.6$~GeV a lot more channels would have to be added to describe the total $e^+e^-\to~{\rm hadrons}$ data. But below 1.6~GeV, our hadronic channels are sufficient to describe the $e^+e^-$-data and hence, we can trust our spectra to describe the physics behind annihilations and decays into hadrons well enough.

As expected, the $B-L$ model is described by a largely reduced set of final states and the gamma spectrum  is dominated by the $\pi^0\gamma, \eta\gamma,$ and $\pi^+\pi^-\pi^0$ channels as well as by the leptonic FSR contribution in the lower energy region. The $\sqrt{s}=0.8$~GeV plot for the $B-L$ model in \cref{fig:DNDX_PHOTON_KM_1GEV} is simply a sum of $\pi^+\pi^-\pi^0$ and $l^+l^-$ for low energies, completely dominated by $\pi^+\pi^-\pi^0$ in the bulk part of the spectrum and characterized by a spectral line from $\pi^0\gamma$ around $x=1$. For $\sqrt{s}\sim m_\phi \sim 1$~GeV, the $KK$ and $\eta\gamma$ channels that are mediated by the $\phi$ meson, come into play with their resonant contributions. The $\eta\gamma$ channel mainly contributes with an additional peak structure around $x=0.7$ due to the final state photon. The $KK$ mode delivers photons from kaon decays as listed in \cref{tab:mesondecays}. Compared to 0.8~GeV and 1.2~GeV where the three-pion channel is completely dominating the bulk part of the spectrum, for 1~GeV several contributions add up to give the total spectral shape. At around 1~GeV, the three-pion channel suffers from a negative interference effect between the $\omega$ and $\phi$ meson contributions. Hence, the $KK$ modes as well as the $\eta\gamma$ contributions give comparable, if not even much larger contributions at 1~GeV as seen in \cref{fig:BR_BL}. Besides, the total spectrum is overall smaller due to the decrease of the $3\pi$ channel.

This can be also seen in the lower left panel of \cref{fig:DNDX_RANGE} where the total gamma spectrum for the $B-L$ model is shown for several center-of-mass energies. Between $0.2-0.6$~GeV, FSR is dominating the total spectrum just like in the kinetic mixing model with an increasing contribution from the $\pi\gamma$ channel for $x\sim 1$. Compared to kinetic mixing where numerous channels contribute, the effect of center-of-mass energy values close to resonances makes a bigger difference and results in the largest total gamma spectrum for 800 MeV among the chosen energy steps. For 1 GeV, we have a reduced spectrum due to the $3\pi$ suppression. The three-pion channels recovers for higher energies and, in addition, contributions from $\eta$ channels, and neutral pions from $KK\pi$ or $\omega\pi\pi$ start to play a role. Hence, we observe a rise of the spectrum until 1.6 GeV. 

\subsection{Positron spectra}

\begin{figure}[t]
  \includegraphics[width=\textwidth]{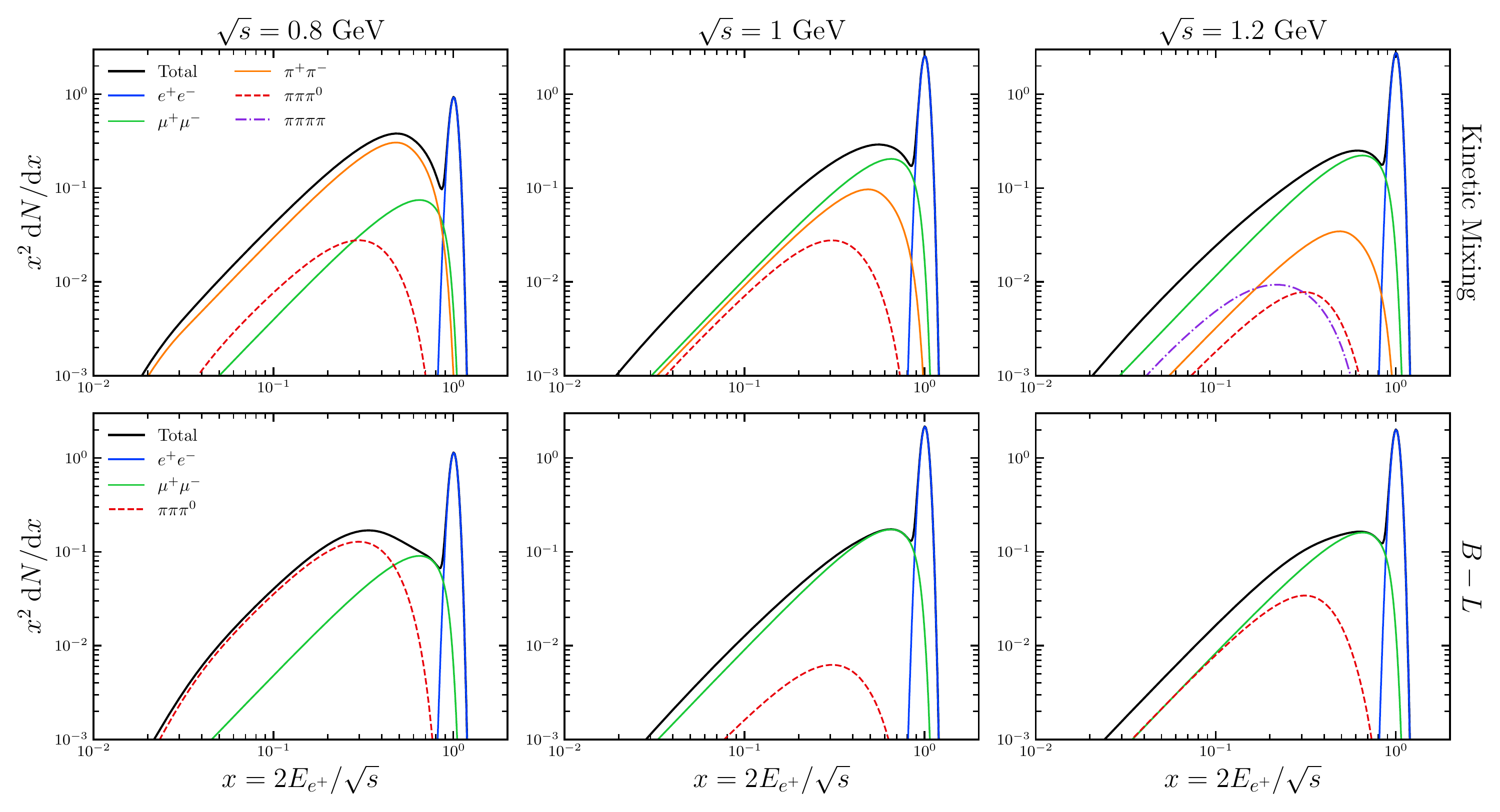}
  \caption{Same as \cref{fig:DNDX_PHOTON_KM_1GEV} but for positrons.}
  \label{fig:DNDX_POSITRON_DETAILED}
\end{figure}

The positron spectra consist of both a line-like spectrum located at $\sqrt{s}/2$ from the direct $e^+e^-$ production and a continuum spectrum from decays of unstable particles. The major ingredient to compute the continuum piece is the $e^+$ energy spectrum from muon decays $\mu^+\to e^+\nu_e \bar{\nu}_\mu$, either directly from the $\mu^+\mu^-$ channel, or through a decay chain that produces charged pions followed by $\pi^+\to \mu^+\nu_\mu \to e^+\nu_e\bar{\nu}_\mu \nu_\mu$. The differential decay probability to obtain a positron with certain energy $E_e$ is commonly described by the Michel decay spectrum~\cite{Michel:1949qe,Fetscher:1986uj} which describes the phase space distribution of charged leptons $l_j$ in $l_i \to l_j \nu_i \bar{\nu}_j$ decays and is often expanded to leading order in $(m_e/m_\mu)$. We use the results beyond leading order in $(m_e/m_\mu)$ as in Ref.~\cite{Coogan:2019qpu} for the $e^+$ spectrum in the muon rest frame $r$. In addition, we can make use of the method for kinematics of a multi-step cascade~\cite{Elor:2015tva} to obtain the analytic form
\begin{align}
  \frac{{\rm d}N_{e^+}}{{\rm d}x_l} & =\frac{1}{2\beta}\int_{x_r^-}^{x_r^+} \frac{{\rm d}x_r}{\sqrt{x_r^2-\mu_r^2}}\frac{{\rm d}N_{e^+}}{{\rm d}x_r} \notag \\
                                    & =-\frac{1}{\beta}\left[\frac23 x^3-\frac32 (1+\frac{\mu_r^2}{4})x^2+\mu_r^2 x\right]_{x_r^-}^{x_r^+}
  \label{eq:muonspec}
\end{align}
for the positron spectrum from muon decays in the lab frame $l$ where we defined the scaleless variables $x_i=2E_i/Q_i$ and $\mu_i=2m_e/Q_i$ with a center-of-mass energy $Q_i$ in the frame $i=r,l$. The bounds of integration are given by
\begin{align}
  x_r^- & ={\rm max}(\mu_r,\gamma^2(x_l-\beta\sqrt{x_l^2-\mu_l^2}))\notag \\
  x_r^+ & ={\rm min}(1+\mu_r^2/4,\gamma^2(x_l+\beta\sqrt{x_l^2-\mu_l^2}))
\end{align}
where the first entry of the max/min function comes from the limits of the ${\rm d}N_{e^+}/{\rm d}x_r$ function itself, or more precisely the kinematic range for the positron energy in the muon rest frame, whereas the second entry is the integration limit for the positron energy in the muon rest frame expressed in lab frame quantities. It is obtained by a change of variables from the angle of $e^+$ with the $z$-axis in the muon rest frame to the positron energy $E_r$. The boost factor is given by $\gamma=E_\mu/m_\mu$ where $E_\mu$ is the muon energy in the lab frame and $\beta=\sqrt{1-1/\gamma^2}$. Note that the dependence on the lab frame positron energy $E_l$ only enters through the bounds of integration. We find that the constant value $x_r^+=1+\mu_r/4$ is picked at approximately $E_l \approx \frac{E_\mu}{2\gamma^2(1+\beta)}$. For a center-of-mass energy of $\sqrt{s}=1$~GeV this corresponds to approximately $x_l=1.1\cdot 10^{-2}$. 

At around $x_r^- \approx 1/(\gamma^2(1-b))$, the lower bound of integration outruns the upper one and no positrons will be found above that value. In the case of $\sqrt{s}=1$~GeV, this happens at approximately $x_r^-\approx 0.99$. As a consequence, the resulting positron spectrum from the $\mu^+\mu^-$ channel starts with a an increasing part coming from the analytic formula above followed by a flattening of the curve above $x\sim 1/(\gamma^2(1-\beta))$. Finally, the available phase space for positrons shrinks down as $E_{e^+}\to E_\mu$ and consequently the spectrum vanishes ${\rm d}N/{\rm d}x \to 0$. For the scaling of the curves in \cref{fig:DNDX_POSITRON_DETAILED}, this translates to a steep increasement, followed by a slope 2 linear part and finally a sharp drop. The $\pi^+\pi^-$ channel shows a similar behaviour since charged pions decay almost entirely into muons. Hence, the spectrum only differs due to shifted boost factors and a smoothed out kink. For higher multiplicity pion final states such as $\pi^+\pi^-\pi^0$ and $4\pi$, or $\eta\gamma\to \pi^+\pi^-\pi^0\gamma$ the characteristic almost box-like shape is getting more round. We can observe in \cref{fig:DNDX_POSITRON_DETAILED} that the sharp drop is shifted to the left for more massive particles in the final states and the height of the curve depends on the branching ratio at a certain energy value. For the kinetic mixing model, the $\pi^+\pi^-$ channel is decreasing while the $\mu^+\mu^-$ channel is increasing with energy. The $\pi^+\pi^-\pi^0$ mode is non-negligible but sub-dominant for all shown energies. For 1.2~GeV, the $\pi^+\pi^-\pi^+\pi^-$ channel is contributing to the overall spectrum as well with a small fraction. 

The situation is even simpler for the $B-L$ model where only $e^+e^-$, $\mu^+\mu^-$, and the $\pi^+\pi^-\pi^0$ channels play a role. As discussed before, for 1~GeV a negative interference of the $\omega$ and $\phi$ contributions to the three-pion channels causes a small branching fraction and hence, the spectrum is completely dominated by leptonic decays. 

The flatness of the spectrum from $\mu^+\mu^-$ can drastically change depending on the center-of-mass energy of the underlying process which determines the muon boost in $\mu^+\mu^-$. For low $\sqrt{s}$ and low boosts, we expect that $\beta$ is small and hence the limiting cases $x_r^\pm \to 1/(\gamma^2(1\pm \beta))$ are very close together resulting in only a small flat part of the spectrum. In the limit of no boost, the spectrum is simply given by the analytic form of \cref{eq:muonspec} with an upper limit $x_r^+=1+\mu_r^2/4$. On the contrary, for very large boosts the box-like shape that is represented by linear slope 2 in our plots will dominate the whole spectrum. In the middle panels of \cref{fig:DNDX_RANGE}, we can see that for lower energy values we still have steep increase followed by a kink and a linear shape. For higher energy values, mostly the linear shape is visible in the plots. For all energies, there is a peak structure from $e^+e^-$ in the final states. For $\sqrt{s}=0.2$~GeV this is even the only visible structure for both the kinetic mixing and $B-L$ model since it is below the $2m_\mu$ threshold to produce muons as well as hadrons.

\subsection{Neutrino Spectra} 

\begin{figure}[t]
  \includegraphics[width=\textwidth]{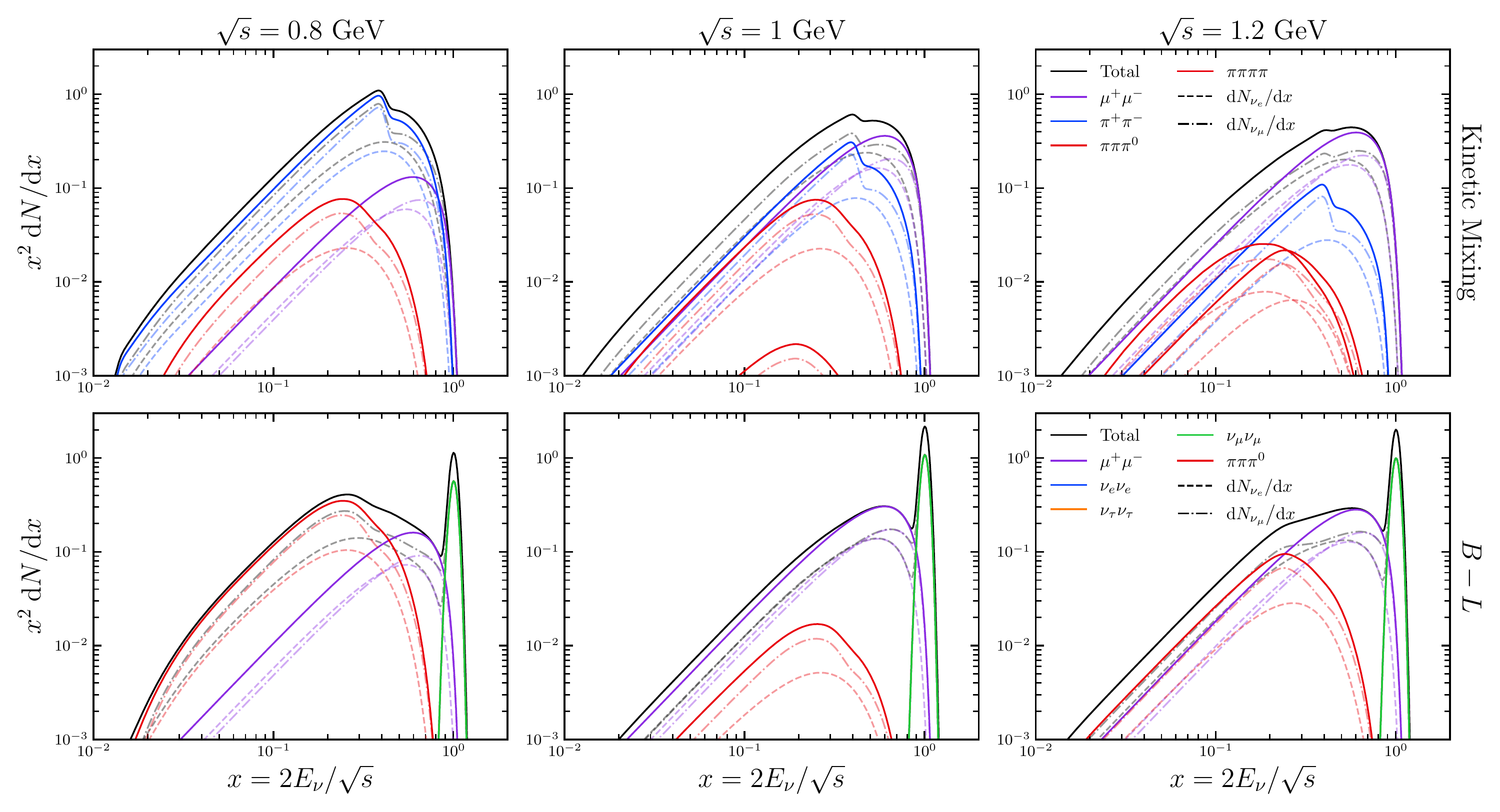}
  \caption{
    Same as \cref{fig:DNDX_PHOTON_KM_1GEV} but for neutrinos. In these plots, the solid lines are the spectra summed over all neutrino species. The dashed lines are the electron-neutrino components and the dash-dotted lines are the muon-neutrino components. The \(\tau\)-neutrino only shows up as a line in the \(B-L\) case, since the final states we consider do not produce \(\tau\)-neutrinos.
  }
  \label{fig:DNDX_NEUTRINO_DETAILED}
\end{figure}

Here we explicitly calculate and show neutrino spectra implemented in \HazmaTwo, which we ignored in previous work~\cite{Coogan:2019qpu,Coogan:2021sjs}. The most important ingredient is the description of how the energy of the neutrinos is distributed in muon decays. The 
neutrino spectra in the muon rest frame is given by
\begin{align}
    {\dv{N_{\mu\to\nu_{e}}}{x}}(x)
    &= 
    \frac{12x^2}{\qty(1-x)R}
    \qty(1 - r^2 - x)^2\\
    \dv{N_{\mu\to\nu_{\mu}}}{x} 
    &= 
    \frac{2x^2}{(1-x)^3 R}
    \qty(1 - r^2 - x)^2
    \qty(3 + r^2(3 -x) - 5x + 2x^2)
\end{align}
where \(r = m_{e}/m_{\mu}\), and \(R = 1 - 8r^2 - 24r^4\log(r) + 8r^6 - r^8\sim 1\). The limits of integration on \(x\) are \(0 \leq x \leq 1-r^2\).
The boosted muon spectra is obtained by computing:
\begin{align}
    {\dv{N_{\mu\to\nu}}{x}}(x,\beta) 
    &= 
    \frac{1}{2\beta}\int^{x_+}_{x_-}
    \frac{\dd{y}}{y}
    {\dv{N_{\mu\to\nu}}{x}}(y)
\end{align}
where \(\)
\(\beta=\sqrt{1-1/\gamma^2}\) is the muon velocity and \(\gamma = E_{\mu}/m_{\mu}\). The values \(x_{\pm} = \gamma^2x\qty(1\pm\beta)\) are the maximum and minimum boosted neutrino energies. The integrals can be computed analytically, and we obtain 
\begin{align}
    \dv{N_{\mu\to\nu_{e}}}{x} 
    &= 
    \frac{1}{\beta R}\qty[\bar{x}_{12}\qty(
        2x_{12}^2 -\qty(3-6r^2)x_{12} + 6r^4 - 2x_{1}x_{2}
    )
    + 6r^4 \log(\frac{1-x_{2}}{1-x_{1}})
    ]
    \\
    \dv{N_{\mu\to\nu_{\mu}}}{x} 
    &= 
    \frac{1}{2\beta R}\bigg{[}
        2r^6\frac{\bar{x}_{12}(1-x_1x_2)}{\qty(1-x_1)^2\qty(1-x_2)^2}
        -6r^4\frac{\bar{x}_{12}}{(1-x_1)(1-x_2)}
        -3r^2 \bar{x}_{12} x_{12}\\
        &\hspace{1.5cm}-\frac{1}{3}\qty(x_1^2\qty(9-4x_1) + x_2^2\qty(9-4x_2))
        -2r^4\qty(3-r^2)\log(\frac{1-x_2}{1-x_1})
    \bigg{]}\notag,
\end{align}
where \(x_{12} \equiv x_{1}+x_{2}\) and \(\bar{x}_{12} \equiv x_2-x_1\). The values \(x_{1}\) and \(x_{2}\) are integration limits when we take into account the integrand is only non-zero if \(0\leq x \leq 1-r^2\). Their values are:
\begin{align}
    x_{1} &= \mathrm{max}\qty(0, \gamma^2x\qty(1-\beta)),&
    x_{2} &= \mathrm{min}\qty(1-r^2, \gamma^2x\qty(1+\beta)).
\end{align}

\begin{figure}
  \centering
  \includegraphics[width=\textwidth]{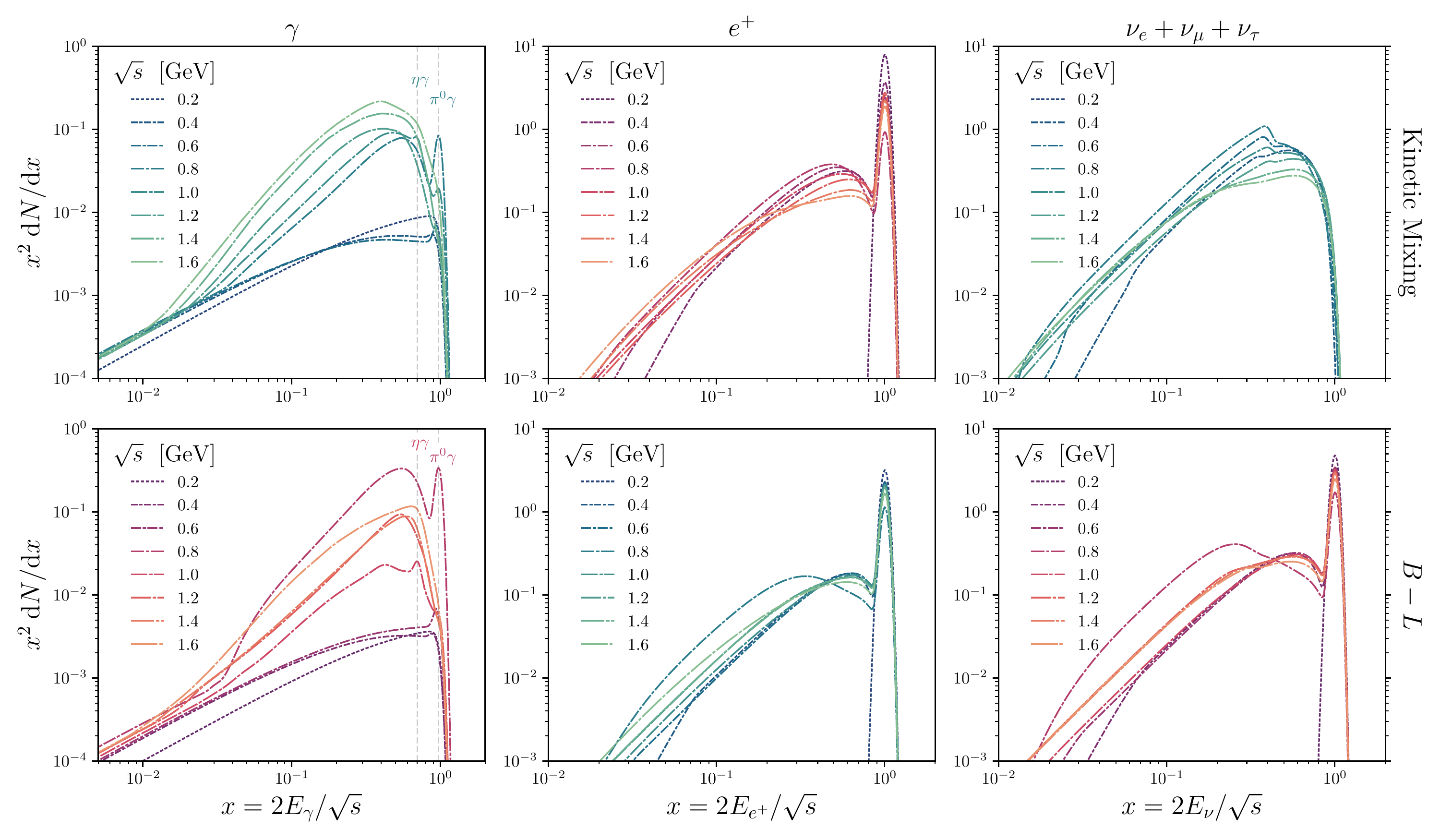}
  \caption{Spectrum from DM annihilations / vector decays for a range of center-of-mass energies. The \textbf{left} plots show the photon spectra for center-of-mass energies between \(200 \ \mathrm{MeV}\) and \(1.6 \ \mathrm{GeV}\). In the \textbf{middle} plots, we should the positron spectra for the same range. The \textbf{right} plots show the neutrino spectra summed over all neutrino flavors. In the \textbf{top} row, we show the spectra assuming kinetic-mixing couplings. In the \textbf{bottom} row, we show the same spectra but assume \(B-L\) couplings. The curves are styled such that the lighter the color and longer the line-length, the higher the energy. The peak structures arise for lines (\(\dv*{N}{x}\sim \delta(x-x_0)\)), which we convolved with a spectral resolution function using a 5\% energy resolution.
  }
  \label{fig:DNDX_RANGE}
\end{figure}

As mentioned before, the production of $e^\pm$ and neutrinos go hand in hand since neutrinos and positrons almost entirely come from charged pion decays $\pi \to \mu\nu \to e3\nu$. Hence the relevant processes for obtaining the $e^+$ spectra as in \cref{fig:DNDX_POSITRON_DETAILED} are also relevant for neutrino production, as can directly be seen in \cref{fig:DNDX_NEUTRINO_DETAILED}. We observe a decreasing $\pi^+\pi^-$ contribution for higher $\sqrt{s}$ whereas the muonic sub-spectrum is becoming more and more dominant for the kinetic mixing model. 

For the $B-L$ model only the $\mu^+\mu^-$ and $3\pi$ channels are playing a role which are equally important for the 0.8~GeV spectrum. For higher energies of 1~GeV and 1.2~GeV, the spectra are mostly driven by $\mu^+\mu^-$. Additionally, there are direct neutrino production modes in the $B-L$ model compared to the kinetic mixing model causing a line-like feature for $x\sim 1$. Except of that $\nu$-line from $\nu_l \bar{\nu}_l$, there is another visible peak structure in the muon neutrino spectrum for the kinetic mixing model directly coming from the charged pion decay $\pi^+ \to \mu^+ \nu_\mu$ in the $\pi^+\pi^-$ channel. The peak mostly contributes significantly to the total spectrum for lower $\sqrt{s}\sim 0.8$~GeV. Note that the peak is not present for the $B-L$ model, that only shows a broader structure due to the $3\pi$ channel. 

\subsection{Comparison to other tools}

\begin{figure}
  \centering
  \includegraphics[width=\textwidth]{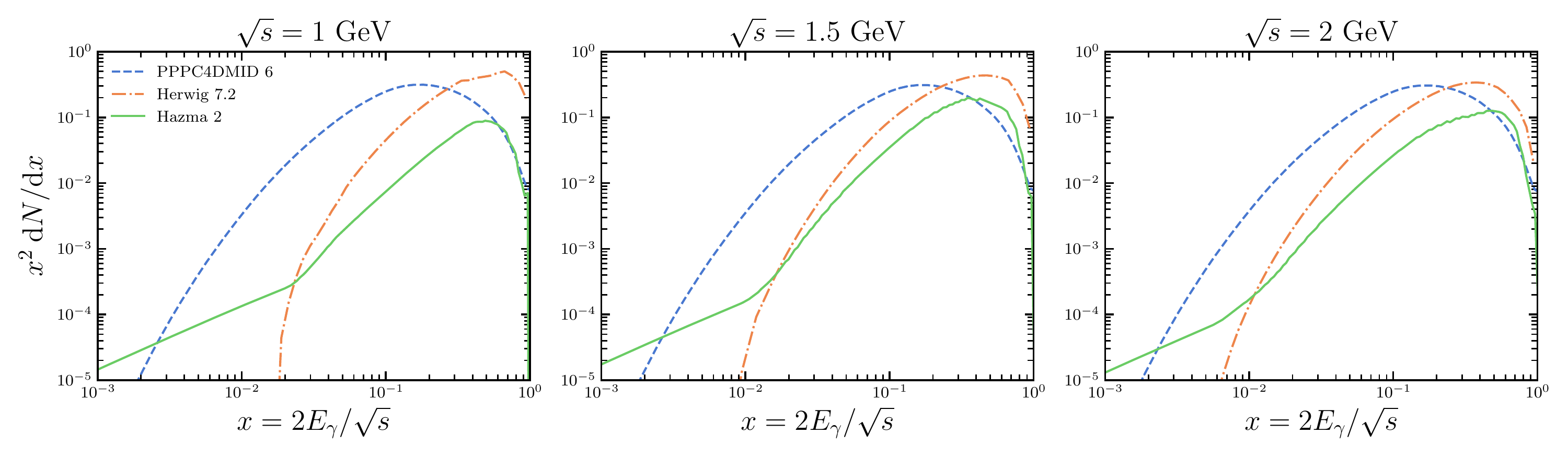}
  \caption{Comparison of the photon spectrum from $\chi\bar{\chi}\to q\bar{q}$ generated using (blue) \textbf{PPPC4DMID v6}, (orange) \textbf{Herwig v7.2} and (green) \HazmaTwo. Note that neither \textbf{PPPC4DMID} nor \textbf{Herwig} are designed to be accurate for these center-of-mass energies.
  }
  \label{fig:DNDX_PHH}
\end{figure}

Available codes for the calculation of photon and positron spectra are not yet able to describe DM annihilation or decay at, and below, $\Lambda_{\rm QCD}$. We choose PPPC4DMID as an example for an inter- and extrapolating tool, and \Herwig\ as a MC tool to generate spectra to compare them to the output of \HazmaTwo\ for the energy values $\sqrt{s}=1, 1.5, 2.0$~GeV in \cref{fig:DNDX_PHH}. To allow for an accurate comparison, we only assume SM-like quark couplings and hence only consider the process $\chi\bar{\chi}\to q\bar{q}$, ignoring leptons.

Unsurprisingly, the PPPC4DMID results do not vary over these low energy values since we expect them to be roughly fixed to its lowest tabulated value of $\sqrt{s}=10$~GeV. In comparison, the \Herwig\ results change with energy since final states are only produced if they are kinematically allowed. Hence, the results from \Herwig\ can roughly mimic the shape of the spectrum qualitatively in the area of where photons from meson decays dominate. Nevertheless, it does not match our \HazmaTwo\ results quantitatively since no accurate description of how to weight several neutral final state configurations such as $\rho\pi^0$ or $\omega\pi^0$ exists. Besides, FSR is only part of the parton shower within MC generators and is ignored at the level of hadrons after hadronization. Hence, photons will only be produced in hadronic decays but not through FSR of charged hadrons in \Herwig. Whereas we are confident that our description for the spectra is accurate up to 1.5~GeV, we expect that the true spectral curve for 2~GeV should lie somewhere in between the \HazmaTwo\ and \Herwig\ result in \cref{fig:DNDX_PHH} since we most likely are missing some additional channels above 1.5 GeV, and \Herwig\ is most likely an overestimation of the true spectrum just as in the lower energy cases before.

\section{Constraints on annihilation and decay rates}


We set conservative constraints utilizing observations of the Milky Way by COMPTEL~\cite{Kappadath:1993}, INTEGRAL~\cite{Bouchet:2011fn}, EGRET~\cite{Strong_2004}, and the Fermi Large Area Telescope (LAT)~\cite{Ackermann:2012pya} with the same procedure described in Ref.~\cite{Coogan:2019qpu}. To briefly summarize, we do not assume a background model, but instead demand the DM signal is consistent with the data. In more detail, we require that the spectral contribution does not exceed the measured gamma-ray flux plus twice the upper error bar in any spectral bin. We adopt the Navarro-Frenk-White profile for the dark matter density profile from Ref.~\cite{Cirelli:2010xx}.

\begin{figure}
  \centering
  \includegraphics[width=\textwidth]{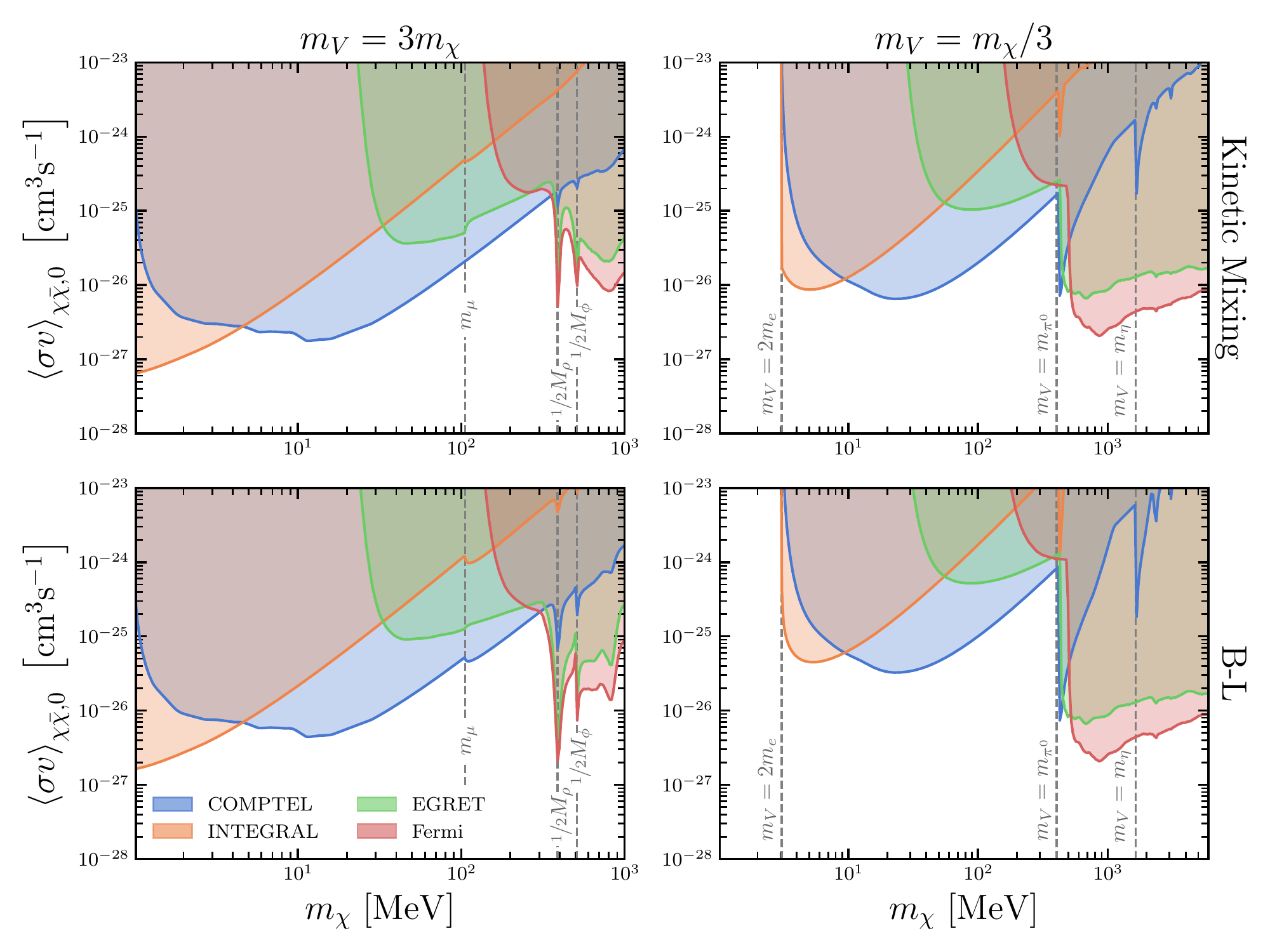}
  \caption{Constraints on the DM velocity-averaged annihilation cross section $\expval{\sigma v}$. In the top and bottom rows, we show results assuming kinetic-mixing and $B-L$ couplings, respectively.
  (\textbf{left}) Results given $m_{V}=3m_{\chi}$. In this case, non-relativistic DM cannot annihilate into $V$. Instead, it must annihilate directly into SM particles.
  (\textbf{right}) Results for $m_{V} = \frac{1}{3}m_{\chi}$, assuming $g_{V\chi\chi}=1$. In this case, the branching ratio $\mathrm{BR}(\chi\bar{\chi}\to VV)$ is $\order{1}$ and photons are produced from the subsequent decay of the vector mediator.}
  \label{fig:constraints_km_bl}
\end{figure}

In \cref{fig:constraints_km_bl} we show the limits we obtain with \HazmaTwo\ on the thermally-averaged self-annihilation cross section times relative velocity for the kinetic mixing and for the $B-L$ models as a function of the dark matter mass $m_\chi$ in the range of validity of our code. The left column assumes that the vector mediator mass $m_V=3 m_\chi$, while the right column takes $m_V=m_\chi/3$, and hence kinematically allows for annihiation directly into mediators. 

The notable features in the plot correspond to a few key kinematic threshold, which we highlight with vertical dashed grey lines. In the right column, the muon threshold produces a slight increase in the limits due to the additional FSR from the muon channel. Additionally the $\rho$ and $\phi$ resonances produce a plethora of additional photon-producing final-state hadrons, and correspondingly more stringent limits for telescopes whose sensitivity covers the relevant photon energies. Conversely, in the case where mediators can be produced in the final state (right column) virtually no photons are produced up to the point where the mediators can decay into electron-positron pairs (left-most vertical dashed line). At larger masses, the two key thresholds then correspond to the production of neutral pions and $\eta$ mesons. 

The constraints illustrate the critical importance of accurately factoring in hadronic resonances for indirect detection in the MeV mass range, as well as the relevance of detailed model-dependent calculations of the relevant hadronic and leptonic final states, both features fully implemented in \HazmaTwo.

\section{Discussion and conclusions}
\label{sec:discussion}

In this work we presented \HazmaTwo\, a tool to close the ``MeV gap'' for theories describing vector-mediated DM annihilation and spin-1 DM decays by merging features of both the \HazmaOne\ and \HerwigDM\ codes. We use the full indirect detection pipeline of \HazmaOne\ in calculating the underlying hard particle physics process including FSR and the energy spectra of photons, positrons and neutrinos as well as ultimately the calculation of limits and predictions for present and future experiments, respectively. From \HerwigDM\ we make use of the description of the processes in the hadronic resonance region, and in particular cross-section formulas that are taken from fits to $e^+e^-$ data. As a result, \HazmaTwo\ constitutes the most accurate and comprehensive tool describing MeV-to-GeV indirect detection for DM models with vector bosons with arbitrary couplings to quarks and leptons.

We have shown that depending on the choice of couplings the branching ratios into hadronic final states can dramatically vary. As two well-motivated and widely discussed example models we have chosen the ``dark photon'' kinetic mixing model and a $B-L$ model. Whereas a wide range of hadronic channels contribute to the total spectra in the dark photon case, the $B-L$ model only couples to a subset of channels. This impacts the size and shape of the spectra. The spectra show distinct features stemming from certain hadronic modes. Especially at lower energies, the $\pi\gamma$ and $\eta\gamma$ channels leave line-like structures on the photon spectra in both models. If one hadronic mode dominates the spectrum as in the case of the $3\pi$ channel in the $B-L$ model, we observed that the photon spectrum changes significantly depending on whether the center-of-mass energy is close to a resonance or not.  We detailed the spectral features for photons, positrons and neutrinos, and tied them all to the relevant hadronic or leptonic final states for the two dark matter models we considered.

We discussed how \HazmaTwo\ compares to other codes when all are tested outside their nominal range of validity. The cautionary tale we derived from this exercise is that quite likely PPPC4DMID and \Herwig\ tend to largely overestimate the photon production for light dark matter compared to what our code indicates, but not at the highest energies, often the most critical to set constraints. Finally, we used \HazmaTwo\ to derive conservative constraints our benchmark DM models using existing gamma-ray data, showing the importance of properly accounting for hadronic resonances. We anticipate \HazmaTwo\ will be a useful tool for leveraging future MeV gamma-ray telescopes to probe the nature of dark matter, and we have made it freely publicly available on GitHub \href{https://github.com/LoganAMorrison/Hazma}{\faGithub}.

\section*{Acknowledgements} 

A.C.\ received funding from the Netherlands eScience Center (grant number ETEC.2019.018) and the Schmidt Futures Foundation. S.P. and L.M. are partly supported by the U.S.\ Department of Energy grant number de-sc0010107. P.R. acknowledges financial support from the Fundação de Amparo à Pesquisa do Estado de São Paulo (FAPESP) under the contract 2020/10004-7. TP is supported by the Deutsche Forschungsgemeinschaft (DFG, German Research Foundation) under grant 396021762 -- TRR~257: \textsl{Particle Physics Phenomenology after the Higgs Discovery} and through Germany's Excellence Strategy EXC 2181/1 - 390900948 (the Heidelberg STRUCTURES Excellence Cluster).

\bibliographystyle{unsrtnat}
\bibliography{main}

\begin{thebibliography}{64}
\providecommand{\natexlab}[1]{#1}
\providecommand{\url}[1]{\texttt{#1}}
\expandafter\ifx\csname urlstyle\endcsname\relax
  \providecommand{\doi}[1]{doi: #1}\else
  \providecommand{\doi}{doi: \begingroup \urlstyle{rm}\Url}\fi

\bibitem[Zyla et~al.(2020)]{ParticleDataGroup:2020ssz}
P.~A. Zyla et~al.
\newblock {Review of Particle Physics}.
\newblock \emph{PTEP}, 2020\penalty0 (8):\penalty0 083C01, 2020.
\newblock \doi{10.1093/ptep/ptaa104}.

\bibitem[Lee and Weinberg(1977)]{Lee:1977ua}
Benjamin~W. Lee and Steven Weinberg.
\newblock {Cosmological Lower Bound on Heavy Neutrino Masses}.
\newblock \emph{Phys. Rev. Lett.}, 39:\penalty0 165--168, 1977.
\newblock \doi{10.1103/PhysRevLett.39.165}.

\bibitem[Profumo et~al.(2019)Profumo, Giani, and Piattella]{Profumo:2019ujg}
Stefano Profumo, Leonardo Giani, and Oliver~F. Piattella.
\newblock {An Introduction to Particle Dark Matter}.
\newblock \emph{Universe}, 5\penalty0 (10):\penalty0 213, 2019.
\newblock \doi{10.3390/universe5100213}.

\bibitem[Coogan et~al.(2020)Coogan, Morrison, and Profumo]{Coogan:2019qpu}
Adam Coogan, Logan Morrison, and Stefano Profumo.
\newblock {Hazma: A Python Toolkit for Studying Indirect Detection of Sub-GeV
  Dark Matter}.
\newblock \emph{JCAP}, 01:\penalty0 056, 2020.
\newblock \doi{10.1088/1475-7516/2020/01/056}.

\bibitem[Mitridate et~al.(2022)Mitridate, Trickle, Zhang, and
  Zurek]{Mitridate:2022tnv}
Andrea Mitridate, Tanner Trickle, Zhengkang Zhang, and Kathryn~M. Zurek.
\newblock {Snowmass White Paper: Light Dark Matter Direct Detection at the
  Interface With Condensed Matter Physics}.
\newblock In \emph{{2022 Snowmass Summer Study}}, 3 2022.

\bibitem[Lehmann and Profumo(2020)]{Lehmann:2020lcv}
Benjamin~V. Lehmann and Stefano Profumo.
\newblock {Cosmology and prospects for sub-MeV dark matter in electron recoil
  experiments}.
\newblock \emph{Phys. Rev. D}, 102\penalty0 (2):\penalty0 023038, 2020.
\newblock \doi{10.1103/PhysRevD.102.023038}.

\bibitem[De~Angelis et~al.(2017)]{e_astrogam}
A.~De~Angelis et~al.
\newblock {Science with e-ASTROGAM: A space mission for MeV\textendash{}GeV
  gamma-ray astrophysics}.
\newblock \emph{Experimental Astronomy}, 44:\penalty0 25, 2017.
\newblock \doi{10.1016/j.jheap.2018.07.001}.

\bibitem[Mallamaci et~al.(2020)]{as_astrogam}
Manuela Mallamaci et~al.
\newblock {All-Sky-ASTROGAM: a MeV Companion for Multimessenger Astrophysics}.
\newblock \emph{PoS}, ICRC2019:\penalty0 579, 2020.
\newblock \doi{10.22323/1.358.0579}.

\bibitem[Orlando et~al.(2021)]{Orlando:2021get}
Elena Orlando et~al.
\newblock {Exploring the MeV Sky with a Combined Coded Mask and Compton
  Telescope: The Galactic Explorer with a Coded Aperture Mask Compton Telescope
  (GECCO)}.
\newblock 12 2021.

\bibitem[Coogan et~al.(2021{\natexlab{a}})Coogan, Moiseev, Morrison, and
  Profumo]{Coogan:2021rez}
Adam Coogan, Alexander Moiseev, Logan Morrison, and Stefano Profumo.
\newblock {Hunting for Dark Matter and New Physics with (a) GECCO}.
\newblock 1 2021{\natexlab{a}}.

\bibitem[{Hunter} et~al.(2010){Hunter}, {Bloser}, {Dion}, {McConnell}, {de
  Nolfo}, {Son}, {Ryan}, and {Stecker}]{adept}
S.~D. {Hunter}, P.~F. {Bloser}, M.~P. {Dion}, M.~L. {McConnell}, G.~A. {de
  Nolfo}, S.~{Son}, J.~M. {Ryan}, and F.~W. {Stecker}.
\newblock {Development of the Advance Energetic Pair Telescope (AdEPT) for
  medium-energy gamma-ray astronomy}.
\newblock In \emph{Space Telescopes and Instrumentation 2010: Ultraviolet to
  Gamma Ray}, volume 7732 of \emph{Proceedings of the SPIE}, page 773221, July
  2010.
\newblock \doi{10.1117/12.857298}.

\bibitem[McEnery et~al.(2019)]{amego}
J.~McEnery et~al.
\newblock {All-sky Medium Energy Gamma-ray Observatory: Exploring the Extreme
  Multimessenger Universe}.
\newblock \emph{Bulletin of the Americal Astronomical Society}, 51:\penalty0
  245, 2019.

\bibitem[Dzhatdoev and Podlesnyi(2019)]{mast}
Timur Dzhatdoev and Egor Podlesnyi.
\newblock {Massive Argon Space Telescope (MAST): A concept of heavy time
  projection chamber for $\gamma$-ray astronomy in the 100 MeV\textendash{}1
  TeV energy range}.
\newblock \emph{Astropart. Phys.}, 112:\penalty0 1--7, 2019.
\newblock \doi{10.1016/j.astropartphys.2019.04.004}.

\bibitem[Tomsick and {\em et al.}(Berlin, 2021)]{Tomsick}
J.A. Tomsick and {\em et al.}
\newblock {The Compton Spectrometer and Imager Project for MeV Astronomy}.
\newblock \emph{PoS(ICRC2021)652, 2021 International Cosmic Ray Conference},
  Berlin, 2021.

\bibitem[Wu et~al.(2014)Wu, Su, Bravar, Chang, Fan, Pohl, and Walter]{pangu}
Xin Wu, Meng Su, Alessandro Bravar, Jin Chang, Yizhong Fan, Martin Pohl, and
  Roland Walter.
\newblock {PANGU: A High Resolution Gamma-ray Space Telescope}.
\newblock \emph{Proc. SPIE Int. Soc. Opt. Eng.}, 9144:\penalty0 91440F, 2014.
\newblock \doi{10.1117/12.2057251}.

\bibitem[Aramaki et~al.(2020)Aramaki, Hansson~Adrian, Karagiorgi, and
  Odaka]{grams}
Tsuguo Aramaki, Per Hansson~Adrian, Georgia Karagiorgi, and Hirokazu Odaka.
\newblock {Dual MeV Gamma-Ray and Dark Matter Observatory - GRAMS Project}.
\newblock \emph{Astropart. Phys.}, 114:\penalty0 107--114, 2020.
\newblock \doi{10.1016/j.astropartphys.2019.07.002}.

\bibitem[Cirelli et~al.(2011)Cirelli, Corcella, Hektor, Hutsi, Kadastik, Panci,
  Raidal, Sala, and Strumia]{Cirelli:2010xx}
Marco Cirelli, Gennaro Corcella, Andi Hektor, Gert Hutsi, Mario Kadastik, Paolo
  Panci, Martti Raidal, Filippo Sala, and Alessandro Strumia.
\newblock {PPPC 4 DM ID: A Poor Particle Physicist Cookbook for Dark Matter
  Indirect Detection}.
\newblock \emph{JCAP}, 03:\penalty0 051, 2011.
\newblock \doi{10.1088/1475-7516/2012/10/E01}.
\newblock [Erratum: JCAP 10, E01 (2012)].

\bibitem[Belanger et~al.(2009)Belanger, Boudjema, Pukhov, and
  Semenov]{Belanger:2008sj}
G.~Belanger, F.~Boudjema, A.~Pukhov, and A.~Semenov.
\newblock {Dark matter direct detection rate in a generic model with micrOMEGAs
  2.2}.
\newblock \emph{Comput. Phys. Commun.}, 180:\penalty0 747--767, 2009.
\newblock \doi{10.1016/j.cpc.2008.11.019}.

\bibitem[Arina et~al.(2021)Arina, Heisig, Maltoni, Massaro, and
  Mattelaer]{Arina:2021gfn}
Chiara Arina, Jan Heisig, Fabio Maltoni, Daniele Massaro, and Olivier
  Mattelaer.
\newblock {Indirect dark-matter detection with MadDM v3.2 -- Lines and Loops}.
\newblock 7 2021.

\bibitem[Bringmann et~al.(2018)Bringmann, Edsj\"o, Gondolo, Ullio, and
  Bergstr\"om]{Bringmann:2018lay}
Torsten Bringmann, Joakim Edsj\"o, Paolo Gondolo, Piero Ullio, and Lars
  Bergstr\"om.
\newblock {DarkSUSY 6 : An Advanced Tool to Compute Dark Matter Properties
  Numerically}.
\newblock \emph{JCAP}, 07:\penalty0 033, 2018.
\newblock \doi{10.1088/1475-7516/2018/07/033}.

\bibitem[Bellm et~al.(2020)]{Bellm:2019zci}
Johannes Bellm et~al.
\newblock {Herwig 7.2 release note}.
\newblock \emph{Eur. Phys. J. C}, 80\penalty0 (5):\penalty0 452, 2020.
\newblock \doi{10.1140/epjc/s10052-020-8011-x}.

\bibitem[Sj\"ostrand(2020)]{Sjostrand:2019zhc}
Torbj\"orn Sj\"ostrand.
\newblock {The PYTHIA Event Generator: Past, Present and Future}.
\newblock \emph{Comput. Phys. Commun.}, 246:\penalty0 106910, 2020.
\newblock \doi{10.1016/j.cpc.2019.106910}.

\bibitem{Amoroso:2018qga}
S.~Amoroso, S.~Caron, A.~Jueid, R.R.~de~Austri and P.~Skands. 
\newblock {Estimating
  {QCD} uncertainties in monte carlo event generators for gamma-ray dark matter
  searches}.
\newblock \emph{JCAP} {\bfseries 2019} (2019) 007.
\newblock \doi{10.1088/1475-7516/2019/05/007}.

\bibitem[Plehn et~al.(2020)Plehn, Reimitz, and Richardson]{Plehn:2019jeo}
Tilman Plehn, Peter Reimitz, and Peter Richardson.
\newblock {Hadronic Footprint of GeV-Mass Dark Matter}.
\newblock \emph{SciPost Phys.}, 8:\penalty0 092, 2020.
\newblock \doi{10.21468/SciPostPhys.8.6.092}.

\bibitem[Reimitz(2021)]{Reimitz:2021wcq}
Peter Reimitz.
\newblock {MeV astronomy with Herwig?}
\newblock \emph{PoS}, TOOLS2020:\penalty0 008, 2021.
\newblock \doi{10.22323/1.392.0008}.

\bibitem[Coogan et~al.(2021{\natexlab{b}})Coogan, Morrison, and
  Profumo]{Coogan:2021sjs}
Adam Coogan, Logan Morrison, and Stefano Profumo.
\newblock {Precision gamma-ray constraints for sub-GeV dark matter models}.
\newblock \emph{JCAP}, 08:\penalty0 044, 2021{\natexlab{b}}.
\newblock \doi{10.1088/1475-7516/2021/08/044}.

\bibitem[Gasser and Leutwyler(1985)]{Gasser:1984gg}
J.~Gasser and H.~Leutwyler.
\newblock {Chiral Perturbation Theory: Expansions in the Mass of the Strange
  Quark}.
\newblock \emph{Nucl. Phys. B}, 250:\penalty0 465--516, 1985.
\newblock \doi{10.1016/0550-3213(85)90492-4}.

\bibitem[Weinberg(1979)]{Weinberg:1978kz}
Steven Weinberg.
\newblock {Phenomenological Lagrangians}.
\newblock \emph{Physica A}, 96\penalty0 (1-2):\penalty0 327--340, 1979.
\newblock \doi{10.1016/0378-4371(79)90223-1}.

\bibitem[Scherer(2003)]{Scherer:2002tk}
Stefan Scherer.
\newblock {Introduction to chiral perturbation theory}.
\newblock \emph{Adv. Nucl. Phys.}, 27:\penalty0 277, 2003.

\bibitem[Bando et~al.(1985{\natexlab{a}})Bando, Kugo, Uehara, Yamawaki, and
  Yanagida]{Bando:1984ej}
M.~Bando, T.~Kugo, S.~Uehara, K.~Yamawaki, and T.~Yanagida.
\newblock {Is rho Meson a Dynamical Gauge Boson of Hidden Local Symmetry?}
\newblock \emph{Phys. Rev. Lett.}, 54:\penalty0 1215, 1985{\natexlab{a}}.
\newblock \doi{10.1103/PhysRevLett.54.1215}.

\bibitem[Kramer et~al.(1984)Kramer, Palmer, and Pinsky]{Kramer:1984cy}
G.~Kramer, William~F. Palmer, and Stephen~S. Pinsky.
\newblock {Testing Chiral Anomalies With Hadronic Currents}.
\newblock \emph{Phys. Rev. D}, 30:\penalty0 89, 1984.
\newblock \doi{10.1103/PhysRevD.30.89}.

\bibitem[Bando et~al.(1988)Bando, Kugo, and Yamawaki]{Bando:1987br}
Masako Bando, Taichiro Kugo, and Koichi Yamawaki.
\newblock {Nonlinear Realization and Hidden Local Symmetries}.
\newblock \emph{Phys. Rept.}, 164:\penalty0 217--314, 1988.
\newblock \doi{10.1016/0370-1573(88)90019-1}.

\bibitem[Bando et~al.(1985{\natexlab{b}})Bando, Kugo, and
  Yamawaki]{Bando:1985rf}
Masako Bando, Taichiro Kugo, and Koichi Yamawaki.
\newblock {On the Vector Mesons as Dynamical Gauge Bosons of Hidden Local
  Symmetries}.
\newblock \emph{Nucl. Phys. B}, 259:\penalty0 493, 1985{\natexlab{b}}.
\newblock \doi{10.1016/0550-3213(85)90647-9}.

\bibitem[Sakurai(1960)]{Sakurai:1960ju}
J.~J. Sakurai.
\newblock {Theory of strong interactions}.
\newblock \emph{Annals Phys.}, 11:\penalty0 1--48, 1960.
\newblock \doi{10.1016/0003-4916(60)90126-3}.

\bibitem[Kroll et~al.(1967)Kroll, Lee, and Zumino]{Kroll:1967it}
N.~M. Kroll, T.~D. Lee, and B.~Zumino.
\newblock {Neutral Vector Mesons and the Hadronic Electromagnetic Current}.
\newblock \emph{Phys. Rev.}, 157:\penalty0 1376--1399, 1967.
\newblock \doi{10.1103/PhysRev.157.1376}.

\bibitem[Lee and Zumino(1967)]{Lee:1967iv}
T.~D. Lee and B.~Zumino.
\newblock {Field Current Identities and Algebra of Fields}.
\newblock \emph{Phys. Rev.}, 163:\penalty0 1667--1681, 1967.
\newblock \doi{10.1103/PhysRev.163.1667}.

\bibitem[Ezhela et~al.(2003)Ezhela, Lugovsky, and Zenin]{Ezhela:2003pp}
V.~V. Ezhela, S.~B. Lugovsky, and O.~V. Zenin.
\newblock {Hadronic part of the muon g-2 estimated on the sigma**2003(tot)(e+
  e- ---\ensuremath{>} hadrons) evaluated data compilation}.
\newblock 12 2003.

\bibitem[Foguel et~al.(2022)Foguel, Reimitz, and Funchal]{Foguel:2022ppx}
Ana~Luisa Foguel, Peter Reimitz, and Renata~Zukanovich Funchal.
\newblock {A robust description of hadronic decays in light vector mediator
  models}.
\newblock \emph{JHEP}, 04:\penalty0 119, 2022.
\newblock \doi{10.1007/JHEP04(2022)119}.

\bibitem[Wess and Zumino(1971)]{Wess:1971yu}
J.~Wess and B.~Zumino.
\newblock {Consequences of anomalous Ward identities}.
\newblock \emph{Phys. Lett. B}, 37:\penalty0 95--97, 1971.
\newblock \doi{10.1016/0370-2693(71)90582-X}.

\bibitem[Witten(1983)]{Witten:1983tw}
Edward Witten.
\newblock {Global Aspects of Current Algebra}.
\newblock \emph{Nucl. Phys. B}, 223:\penalty0 422--432, 1983.
\newblock \doi{10.1016/0550-3213(83)90063-9}.

\bibitem[Fujiwara et~al.(1985)Fujiwara, Kugo, Terao, Uehara, and
  Yamawaki]{Fujiwara:1984mp}
Takanori Fujiwara, Taichiro Kugo, Haruhiko Terao, Shozo Uehara, and Koichi
  Yamawaki.
\newblock {Nonabelian Anomaly and Vector Mesons as Dynamical Gauge Bosons of
  Hidden Local Symmetries}.
\newblock \emph{Prog. Theor. Phys.}, 73:\penalty0 926, 1985.
\newblock \doi{10.1143/PTP.73.926}.

\bibitem[Czyz et~al.(2010)Czyz, Grzelinska, and Kuhn]{Czyz:2010hj}
Henryk Czyz, Agnieszka Grzelinska, and Johann~H. Kuhn.
\newblock {Narrow resonances studies with the radiative return method}.
\newblock \emph{Phys. Rev. D}, 81:\penalty0 094014, 2010.
\newblock \doi{10.1103/PhysRevD.81.094014}.

\bibitem[Czy\.z et~al.(2018)Czy\.z, Kisza, and Tracz]{Czyz:2017veo}
Henryk Czy\.z, Patrycja Kisza, and Szymon Tracz.
\newblock {Modeling interactions of photons with pseudoscalar and vector
  mesons}.
\newblock \emph{Phys. Rev. D}, 97\penalty0 (1):\penalty0 016006, 2018.
\newblock \doi{10.1103/PhysRevD.97.016006}.

\bibitem[Davier et~al.(2011)Davier, Hoecker, Malaescu, and
  Zhang]{Davier:2010nc}
Michel Davier, Andreas Hoecker, Bogdan Malaescu, and Zhiqing Zhang.
\newblock {Reevaluation of the Hadronic Contributions to the Muon g-2 and to
  alpha(MZ)}.
\newblock \emph{Eur. Phys. J. C}, 71:\penalty0 1515, 2011.
\newblock \doi{10.1140/epjc/s10052-012-1874-8}.
\newblock [Erratum: Eur.Phys.J.C 72, 1874 (2012)].

\bibitem[Czyz et~al.(2006)Czyz, Grzelinska, Kuhn, and Rodrigo]{Czyz:2005as}
Henryk Czyz, Agnieszka Grzelinska, Johann~H. Kuhn, and German Rodrigo.
\newblock {Electron-positron annihilation into three pions and the radiative
  return}.
\newblock \emph{Eur. Phys. J. C}, 47:\penalty0 617--624, 2006.
\newblock \doi{10.1140/epjc/s2006-02614-7}.

\bibitem[Czyz et~al.(2008)Czyz, Kuhn, and Wapienik]{Czyz:2008kw}
Henryk Czyz, Johann~H. Kuhn, and Agnieszka Wapienik.
\newblock {Four-pion production in tau decays and e+e- annihilation: An
  Update}.
\newblock \emph{Phys. Rev. D}, 77:\penalty0 114005, 2008.
\newblock \doi{10.1103/PhysRevD.77.114005}.

\bibitem[Aubert et~al.(2007)]{BaBar:2007qju}
Bernard Aubert et~al.
\newblock {The e+ e- ---\ensuremath{>} 2(pi+ pi-) pi0, 2(pi+ pi-) eta, K+ K-
  pi+ pi- pi0 and K+ K- pi+ pi- eta Cross Sections Measured with Initial-State
  Radiation}.
\newblock \emph{Phys. Rev. D}, 76:\penalty0 092005, 2007.
\newblock \doi{10.1103/PhysRevD.76.092005}.
\newblock [Erratum: Phys.Rev.D 77, 119902 (2008)].

\bibitem[Lees et~al.(2018)]{BaBar:2018rkc}
J.~P. Lees et~al.
\newblock {Study of the reactions $e^+e^-\to\pi^+\pi^-\pi^0\pi^0\pi^0\gamma$
  and $\pi^+\pi^-\pi^0\pi^0\eta\gamma$ at center-of-mass energies from
  threshold to 4.35 GeV using initial-state radiation}.
\newblock \emph{Phys. Rev. D}, 98\penalty0 (11):\penalty0 112015, 2018.
\newblock \doi{10.1103/PhysRevD.98.112015}.

\bibitem[Christy et~al.(2022)Christy, Kumar, and Rajaraman]{Christy:2022pvy}
J.~G. Christy, Jason Kumar, and Arvind Rajaraman.
\newblock {Indirect Detection of Low-mass Dark Matter Through the $\pi^0$ and
  $\eta$ Windows}.
\newblock 5 2022.

\bibitem[Low(1958)]{Low:1958sn}
F.~E. Low.
\newblock {Bremsstrahlung of very low-energy quanta in elementary particle
  collisions}.
\newblock \emph{Phys. Rev.}, 110:\penalty0 974--977, 1958.
\newblock \doi{10.1103/PhysRev.110.974}.

\bibitem[Burnett and Kroll(1968)]{Burnett:1967km}
T.~H. Burnett and Norman~M. Kroll.
\newblock {Extension of the low soft photon theorem}.
\newblock \emph{Phys. Rev. Lett.}, 20:\penalty0 86, 1968.
\newblock \doi{10.1103/PhysRevLett.20.86}.

\bibitem[Michel(1950)]{Michel:1949qe}
L.~Michel.
\newblock {Interaction between four half spin particles and the decay of the
  $\mu$ meson}.
\newblock \emph{Proc. Phys. Soc. A}, 63:\penalty0 514--531, 1950.
\newblock \doi{10.1088/0370-1298/63/5/311}.

\bibitem[Fetscher et~al.(1986)Fetscher, Gerber, and Johnson]{Fetscher:1986uj}
W.~Fetscher, H.~J. Gerber, and K.~F. Johnson.
\newblock {Muon Decay: Complete Determination of the Interaction and Comparison
  with the Standard Model}.
\newblock \emph{Phys. Lett. B}, 173:\penalty0 102--106, 1986.
\newblock \doi{10.1016/0370-2693(86)91239-6}.

\bibitem[Elor et~al.(2015)Elor, Rodd, and Slatyer]{Elor:2015tva}
Gilly Elor, Nicholas~L. Rodd, and Tracy~R. Slatyer.
\newblock {Multistep cascade annihilations of dark matter and the Galactic
  Center excess}.
\newblock \emph{Phys. Rev. D}, 91:\penalty0 103531, 2015.
\newblock \doi{10.1103/PhysRevD.91.103531}.

\bibitem[Kappadath(1993)]{Kappadath:1993}
S.~Cheenu Kappadath.
\newblock \emph{Measurement of the Cosmic Diffuse Gamma-Ray Spectrum from 800
  keV to 30 MeV}.
\newblock PhD thesis, University of New Hampshire, May 1993.

\bibitem[Bouchet et~al.(2011)Bouchet, Strong, Porter, Moskalenko, Jourdain, and
  Roques]{Bouchet:2011fn}
Laurent Bouchet, Andrew~W. Strong, Troy~A. Porter, Igor~V. Moskalenko,
  Elisabeth Jourdain, and Jean-Pierre Roques.
\newblock {Diffuse emission measurement with INTEGRAL/SPI as indirect probe of
  cosmic-ray electrons and positrons}.
\newblock \emph{Astrophys. J.}, 739:\penalty0 29, 2011.
\newblock \doi{10.1088/0004-637X/739/1/29}.

\bibitem[Strong et~al.(2004)Strong, Moskalenko, and Reimer]{Strong_2004}
Andrew~W. Strong, Igor~V. Moskalenko, and Olaf Reimer.
\newblock Diffuse galactic continuum gamma rays: A model compatible with
  {EGRET} data and cosmic-ray measurements.
\newblock \emph{The Astrophysical Journal}, 613\penalty0 (2):\penalty0
  962--976, oct 2004.
\newblock \doi{10.1086/423193}.
\newblock URL \url{https://doi.org/10.1086%2F423193}.

\bibitem[Ackermann et~al.(2012)]{Ackermann:2012pya}
M.~Ackermann et~al.
\newblock {Fermi-LAT Observations of the Diffuse Gamma-Ray Emission:
  Implications for Cosmic Rays and the Interstellar Medium}.
\newblock \emph{Astrophys. J.}, 750:\penalty0 3, 2012.
\newblock \doi{10.1088/0004-637X/750/1/3}.

\bibitem[Achasov et~al.(2016)]{SND:2016drm}
M.~N. Achasov et~al.
\newblock {Study of the reaction $e^+e^- \to \pi^0\gamma$ with the SND detector
  at the VEPP-2M collider}.
\newblock \emph{Phys. Rev. D}, 93\penalty0 (9):\penalty0 092001, 2016.
\newblock \doi{10.1103/PhysRevD.93.092001}.

\bibitem[Akhmetshin et~al.(2005)]{Akhmetshin:2004gw}
R.~R. Akhmetshin et~al.
\newblock {Study of the processes {$e^{+} e^{-} \to \eta\gamma, \pi^{0}\gamma
  3\gamma$} in the c.m. energy range 600-MeV to 1380-MeV at CMD-2}.
\newblock \emph{Phys. Lett.}, B605:\penalty0 26--36, 2005.
\newblock \doi{10.1016/j.physletb.2004.11.020}.

\bibitem[Lai et~al.(2007)]{NA48:2006xvd}
A.~Lai et~al.
\newblock {Measurement of K0(mu3) form factors}.
\newblock \emph{Phys. Lett. B}, 647:\penalty0 341--350, 2007.
\newblock \doi{10.1016/j.physletb.2007.02.039}.

\bibitem[Anastasi et~al.(2016)Anastasi, Babusci, Bencivenni, Berlowski, Bloise,
  Bossi, Branchini, Budano, Caldeira~Balkest{\aa}hl, Cao,
  et~al.]{anastasi2016precision}
A~Anastasi, D~Babusci, G~Bencivenni, M~Berlowski, C~Bloise, F~Bossi,
  P~Branchini, A~Budano, L~Caldeira~Balkest{\aa}hl, Bo~Cao, et~al.
\newblock Precision measurement of the $\eta$→ $\pi$+ $\pi$- $\pi$ 0 dalitz
  plot distribution with the kloe detector.
\newblock \emph{Journal of High Energy Physics}, 2016\penalty0 (5):\penalty0
  1--21, 2016.

\bibitem[Prakhov et~al.(2018)Prakhov, Abt, Achenbach, Adlarson, Afzal,
  Aguar-Bartolom{\'e}, Ahmed, Ahrens, Annand, Arends, et~al.]{prakhov2018high}
S~Prakhov, S~Abt, P~Achenbach, P~Adlarson, F~Afzal, P~Aguar-Bartolom{\'e},
  Z~Ahmed, J~Ahrens, JRM Annand, HJ~Arends, et~al.
\newblock High-statistics measurement of the $\eta$→ 3 $\pi$ 0 decay at the
  mainz microtron.
\newblock \emph{Physical Review C}, 97\penalty0 (6):\penalty0 065203, 2018.

\bibitem[Adlarson et~al.(2017)]{WASA-at-COSY:2016hfo}
P.~Adlarson et~al.
\newblock {Measurement of the $\omega \to \pi^+ \pi^- \pi^0$ Dalitz plot
  distribution}.
\newblock \emph{Phys. Lett. B}, 770:\penalty0 418--425, 2017.
\newblock \doi{10.1016/j.physletb.2017.03.050}.

\bibitem[Ablikim et~al.(2015)]{BESIII:2015fid}
M.~Ablikim et~al.
\newblock {Measurement of the Matrix Elements for the Decays $\eta \rightarrow
  \pi^{+}\pi^{-}\pi^0$ and $\eta/\eta^{\prime}\rightarrow\pi^0\pi^0\pi^0$}.
\newblock \emph{Phys. Rev. D}, 92:\penalty0 012014, 2015.
\newblock \doi{10.1103/PhysRevD.92.012014}.

\end{thebibliography}

\appendix

\clearpage

\section{New form factor}
\label{app:formfactor}
The hadronic current of the $\pi\gamma$ as given by \cref{eq:pigamma} is a
simplified version of the parametrization of~\cite{Czyz:2017veo}. The advantage
of the form factor as in~\cite{Czyz:2017veo} compared to,
\textit{i.e.}~\cite{SND:2016drm}, is that the low-energy ChPT results of
\cref{eq:JChPT} is recovered for $\hat{s}\to 0$. Since a photon is part of
the final state configuration, isospin assumptions do not apply for this channel
and all vector meson resonances are assumed to play a role. The channel can be
better described by the $\rho\omega\pi$ and $\rho\phi\pi$ vertices arising from
\cref{eq:XXP} where one outgoing vector meson is mixing with a photon through
the mixing term of \cref{eq:XPP}. In \cref{tab:pigamma}, we present the
results of the fit including the fit uncertainties. The value of the pion decay
constant has been set to $f_\pi=0.09266$~GeV

\begin{table}[h]
  \def\arraystretch{1.2}
  \begin{center}
    \begin{tabular}{|c|c|}
      \hline
      Parameter  & Fit Value                                  \\
      \hline
      $a$        & $0.00759(25)$                              \\
      $a_\rho$   & 1 (fixed)                                  \\
      $a_\omega$ & 0.885(32)                                  \\
      $a_\phi$   & -0.0646(22)                                \\
      \hline
      \multicolumn{2}{|c|}{$\chi^2 \mathrm{/n.d.f.} = 1.049$} \\
      \hline
    \end{tabular}
  \end{center}
  \caption{Fit values for $\pi\gamma$ with fit uncertainties in brackets.}
  \label{tab:pigamma}
\end{table}

\begin{figure}[htb]
  \begin{center}
    \includegraphics[width=0.6\textwidth]{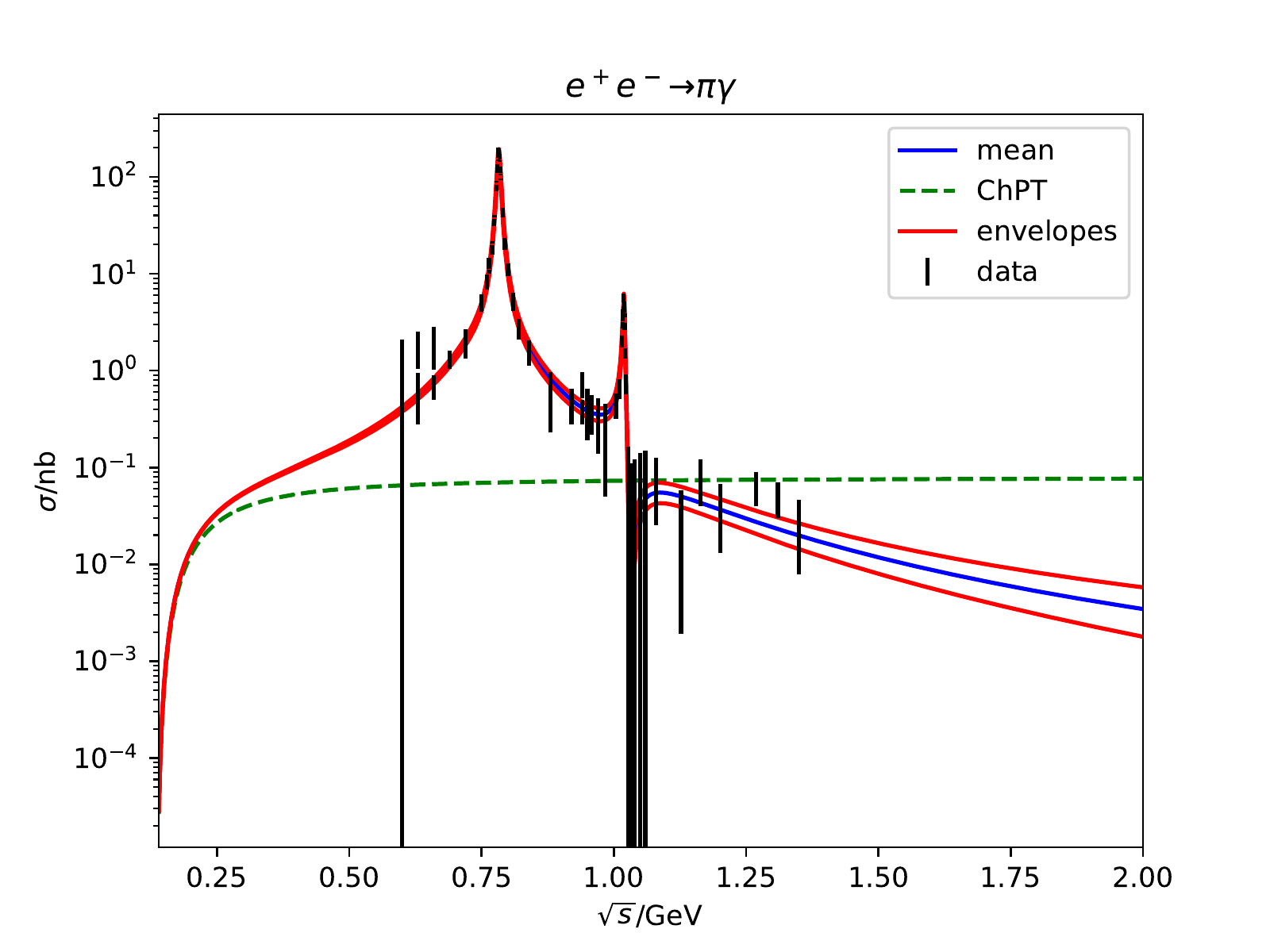}
  \end{center}
  \vglue -0.8cm
  \caption{ Cross-section fit to data for $e^+e^-\to \pi\gamma$. The fit curve
    (blue) is following the data closely in the vector meson resonance region in
    comparison to the ChPT results (green). Both fit and ChPT lines merge in the
    low-energy limit. The error envelopes (red) show the maximal and minimal
    curves we obtain by varying the fit parameters around the fit uncertainties.}
  \label{fig:fitpigamma}
\end{figure}

The $e^+e^-$ annihilation data for the fit has been taken
from~\cite{SND:2016drm,Akhmetshin:2004gw}. As shown in
\cref{fig:fitpigamma}, the ChPT result is obtained in the low-energy limit
and the effects of the vector resonances of $\omega$ and $\phi$ are accurately
described with the parametrization. This also explains the low uncertainties
through fit uncertainties in the low-energy limit as seen in
\cref{fig:fitpigamma} since the ChPT limit is independent of the VMD
parametrization and hence, no uncertainties can be propagated in that limit.

\section{Widths and Cross Sections From Form Factors}

Here we discuss the how to obtain widths and cross-sections from the hadronic
current. Let the hadronic current for some final state \(\cH\) be \(J_{\cH}^{\mu}\).
The matrix elements for dark matter annihilation and vector decays are
given by:
\begin{align}
  \cM_{\bar{\chi}\chi\to\cH} & = \chi_{\mu}\frac{i}{s-M_{V}^{2}+i M_{V}\Gamma_{V}}J_{\cH}^{\mu} \\
  \cM_{V\to\cH}              & = \epsilon_{\mu}(P) J_{\cH}^{\mu}
\end{align}
with \(\chi_{\mu} = \bar{v}(p_2)\qty(ig_{V\chi}\gamma_{\mu})u(p_1)\).
For the moment, we focus on the hadronic current. If we square the hadronic
current and integrate over phase space, the result is of the form
\begin{align}
  \cJ^{\mu\nu} & = \int\dd{\Pi}_{\mathrm{LIPS}} J_{\cH}^{\mu}\bar{J}_{\cH}^{\nu} = \cJ(s)\qty(P_{\mu}P_{\nu} - g_{\mu\nu}P^2)
\end{align}
where \(P = p_{1} + p_{2}\) is the center-of-mass momentum. The last result is
required by gauge-invariance of the hadronic current. Contracting both sides
with the metric tensor, the integrated current can be written as:
\begin{align}
  \cJ(s) & = -\frac{1}{3s}g_{\mu\nu}\cJ^{\mu\nu}
\end{align}
Given this result, we can write the annihilation cross section and decay width as:
\begin{align}
  \sigma_{\bar{\chi}\chi\to\cH} & = \frac{g^{2}_{V\chi\chi}(s+2m_{\chi}^{2})}{\sqrt{s-4m_{\chi}^2}\qty(\qty(s-M_V^2)^2 + M_V^2\Gamma_V^2)}\frac{\sqrt{s}}{2}\cJ(s)
  \\
  \Gamma_{V\to\cH}              & = \frac{\sqrt{s}}{2}\cJ(s)
\end{align}
where \(s = P^2\) is the square center-of-mass energy. For the vector decay, \(s
= M_{V}^2\). However, writing the results in the above form allows us to reuse
results. In addition, we can see that, as long as \(m_{\chi} > M_{V}\), in which
case the \(\bar{\chi}\chi\to VV\) is in inaccessible for small dark matter
velocities, the branching ratios for vector decays are identical to dark matter
annihilations.

\end{document}